\documentclass[10pt,twocolumn,letterpaper]{article}

\usepackage[pagenumbers]{iccv} %
\usepackage{caption}

\usepackage{booktabs} %
\usepackage{graphicx}
\usepackage{amsmath}
\usepackage{booktabs}
\usepackage{multirow}
\usepackage{colortbl}
\usepackage{makecell}
\usepackage{diagbox}
\usepackage{xcolor}
\usepackage{enumitem}

\definecolor{iccvblue}{rgb}{0.21,0.49,0.74}
\usepackage[pagebackref,breaklinks,colorlinks,allcolors=iccvblue]{hyperref}

\def\paperID{3659} %
\def\confName{ICCV}
\def\confYear{2025}

\title{\vspace{-8mm}GENMO: A \underline{GEN}eralist Model for Human \underline{MO}tion\vspace{-3mm}}

\author{
Jiefeng Li \hspace{.5em} Jinkun Cao \hspace{.5em} Haotian Zhang \hspace{.5em} Davis Rempe \hspace{.5em} Jan Kautz \hspace{.5em} Umar Iqbal \hspace{.5em} Ye Yuan \\[1mm]
NVIDIA \\
{ \url{https://research.nvidia.com/labs/dair/genmo}} \\
}

\begin{document}
\twocolumn[{%
\renewcommand\twocolumn[1][]{#1}%
\maketitle
\begin{center}
    \vspace{-5mm}
    \centering
    \captionsetup{type=figure}
    \includegraphics[width=\textwidth]{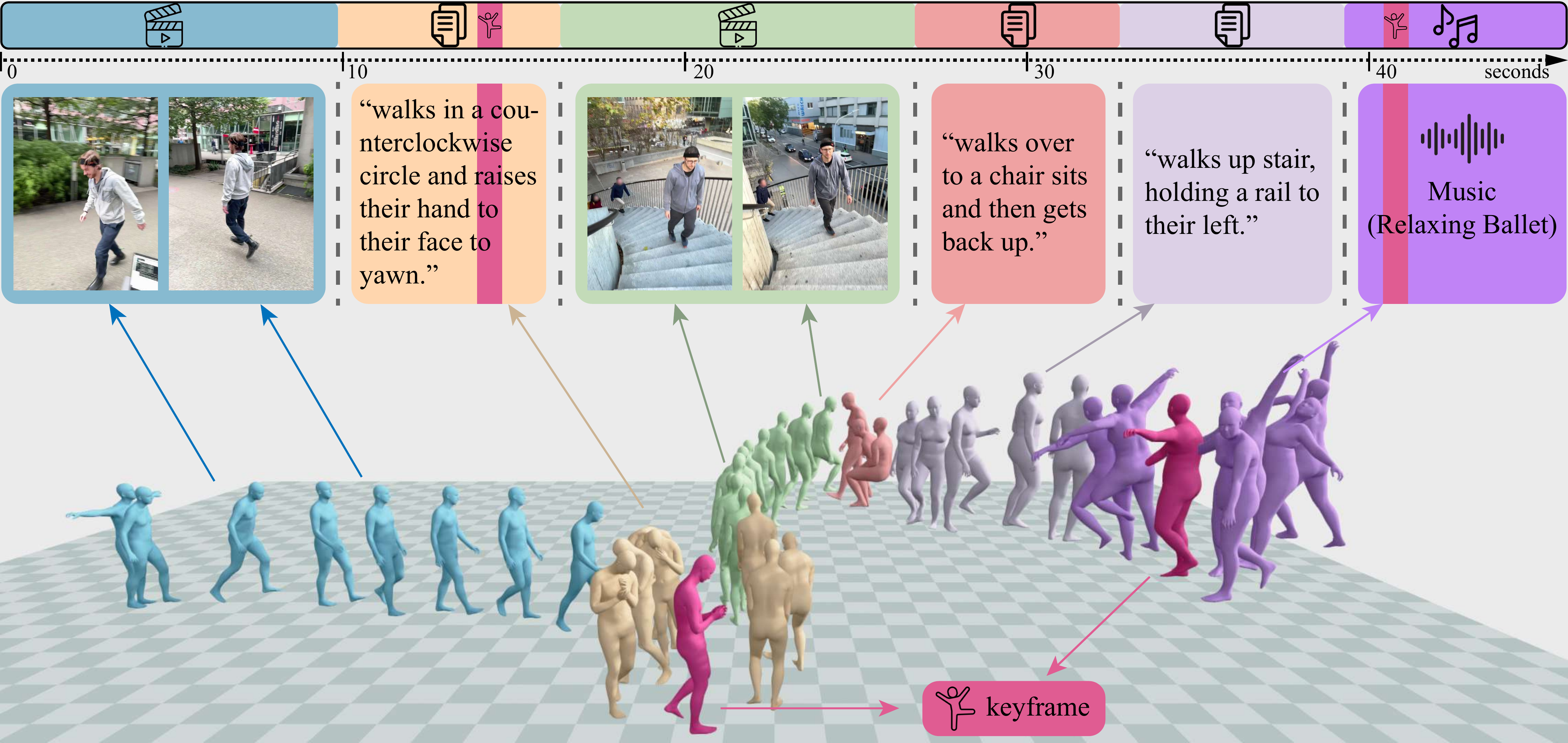}
    \caption{GENMO unifies human motion estimation and generation in a single framework and supports diverse conditioning signals including monocular videos, 2D keypoints, text descriptions, music, and 3D keyframes. GENMO can estimate accurate global human motion from videos with dynamic cameras and seamlessly handles arbitrary combinations and lengths of conditioning signals while generating smooth transitions between them. All of this is achieved in a single feedforward diffusion pass without complex post-processing.}
    \label{fig:teaser}
\end{center}%
}]

\begin{abstract}
Human motion modeling traditionally separates motion generation and estimation into distinct tasks with specialized models. Motion generation models focus on creating diverse, realistic motions from inputs like text, audio, or keyframes, while motion estimation models aim to reconstruct accurate motion trajectories from observations like videos. Despite sharing underlying representations of temporal dynamics and kinematics, this separation limits knowledge transfer between tasks and requires maintaining separate models. We present GENMO, a unified Generalist Model for Human Motion that bridges motion estimation and generation in a single framework. Our key insight is to reformulate motion estimation as constrained motion generation, where the output motion must precisely satisfy observed conditioning signals.  Leveraging the synergy between regression and diffusion, GENMO achieves accurate global motion estimation while enabling diverse motion generation. We also introduce an estimation-guided training objective that exploits in-the-wild videos with 2D annotations and text descriptions to enhance generative diversity. Furthermore, our novel architecture handles variable-length motions and mixed multimodal conditions (text, audio, video) at different time intervals, offering flexible control. This unified approach creates synergistic benefits: generative priors improve estimated motions under challenging conditions like occlusions, while diverse video data enhances generation capabilities. Extensive experiments demonstrate GENMO's effectiveness as a generalist framework that successfully handles multiple human motion tasks within a single model.
\end{abstract}

\section{Introduction}

Human motion modeling is a longstanding topic in computer vision and graphics, with applications in gaming, animation, and 3D content creation. These creative applications typically require precise and intuitive user control. Consider a scenario where a user aims to generate motion sequences integrating multiple modalities: starting from a video clip, transitioning to follow textual descriptions, syncing with audio cues, and aligning with another video, all while providing fine-grained control via user-defined keyframes. Such sequences must precisely replicate observed human movements, reflect intended actions described by text or music, and adhere consistently to specified keyframes. While recent advances have made significant progress in individual tasks, achieving such precision and flexibility across multiple modalities remains challenging. Specifically, motion estimation from videos typically involves deterministic predictions focused on accuracy, whereas text/music-to-motion generation requires diversity to all possible motions. Consequently, these tasks are usually treated independently despite sharing common representations like temporal dynamics and kinematic structures. This separation limits cross-task knowledge transfer and requires maintaining distinct models.

Recent studies have revealed the synergistic relationship between motion estimation and generation tasks. Generative models~\cite{rempe2021humor,he2022nemf, tevet2023human} have provided robust priors for motion estimation, particularly in challenging scenarios such as world-space estimation~\cite{bogo2016keep,ye2023slahmr,kocabas2024pace,li2024coin}. Conversely, leveraging large-scale video data for estimation has enhanced the realism of generative models by enriching their learned motion distributions~\cite{lin2023motionx}. This motivates developing a unified generalist model capable of handling both tasks concurrently across multiple modalities. However, developing such a framework presents significant challenges due to the contrasting objectives of these tasks: generation requires producing diverse and plausible outputs from abstract inputs like text or audio, while estimation demands precise motion reconstruction from concrete observations such as videos and keypoints. Creating a unified architecture that effectively balances diverse generation with accurate reconstruction while leveraging shared representations remains a complex challenge.

To address these issues, we propose GENMO, a Generalist Model for Human Motion that unifies estimation and generation within a single framework. We formulate motion estimation as constrained motion generation adhering to observed signals. This unification yields synergistic benefits: generative priors enhance plausibility in challenging estimation scenarios (e.g., occlusions), while diverse video data enrich generative diversity without requiring ground-truth 3D annotations.

GENMO is built upon a diffusion model framework incorporating a novel dual-mode training paradigm: (1) \textit{estimation mode}, where we feed the GENMO diffusion denoiser with zero-initialized noise and the largest diffusion timestep, forcing the model to produce maximum likelihood estimation (MLE) of the motion based on the conditional signals; (2) \textit{generation mode}, follows traditional diffusion training by sampling noisy motions and timesteps according to a predefined schedule, enabling the model to learn rich generative distributions from the conditioning signals. This dual-mode approach allows GENMO to excel at both precise estimation and diverse generation tasks. We further enhance the framework with an estimation-guided training objective that effectively leverages in-the-wild videos with 2D annotations, substantially expanding the model's generative capabilities. Furthermore, our architectural innovations enable the processing of variable-length motion sequences and seamlessly integrate arbitrary combinations of multi-modal conditioning signals at different time intervals, as demonstrated in Fig.~\ref{fig:teaser}. Notably, GENMO generates multi-conditioned motions in a single feedforward diffusion pass, without requiring complex post-processing steps.

Through extensive empirical evaluation, we demonstrate GENMO's capabilities across a comprehensive suite of tasks encompassing both global and local motion estimation, as well as diverse motion generation tasks including music-to-dance synthesis, text-to-motion generation, and motion-inbetweening. Our experimental results establish that GENMO achieves state-of-the-art performance across various tasks (global motion estimation, local motion estimation, and music-to-dance generation), validating its efficacy as a unified generalist framework for human motion modeling.

Our contributions are summarized as follows:

\begin{itemize}[left=5pt, topsep=5pt]
    \item We propose GENMO, the first generalist model unifying state-of-the-art global motion estimation with flexible human motion generation conditioned on videos, music, text, 2D keypoints, and 3D keyframes.
    \item Our architecture design supports seamless generation of variable-length motions conditioned on arbitrary numbers and combinations of multimodal inputs without complex post-processing.
    \item We propose a novel dual-mode training paradigm to explore the synergy between regression and diffusion, and introduce an estimation-guided training objective that enables effective training on in-the-wild videos.
    \item We demonstrate bidirectional benefits: generative priors improve estimation under challenging conditions like occlusions; conversely, diverse video data enhances generative expressiveness.
\end{itemize}
\section{Related Work}

\subsection{Human Motion Generation}

Human motion generation has progressed significantly in recent years~\cite{Habibie2017ARV,BarsoumCVPRW2018,Henter2020,valleperez2021transflower,chopin21_wgan,rempe2021humor,zhang2022motiondiffuse,he2022nemf,cervantes2022implicit,guo2023momask,zhu2023human,pinyoanuntapong2024mmm,Shiobara21_wgan,xu2023actformer,tevet2023human, zhang2023remodiffuse, pinyoanuntapong2024mmm,cohan2024flexible, motionlcm} leveraging a variety of conditioning signals such as
text~\cite{Ghosh_2021_ICCV,chuan2022tm2t,chen2023mld,jin2023act}, 
actions~\cite{chuan2020action2motion}, 
speech~\cite{zhu2023taming,alexanderson2023listen}, 
music~\cite{tang2018_dance,li2021ai,valleperez2021transflower,sun2022_dancing,siyao2022bailando,tseng2023edge},
and scenes/objects~\cite{wang2022humanise,hassan2021stochastic,kulkarni2023nifty, zhang2024force, yi2025generating}. Recently, multimodal motion generation has also gained attention~\cite{Zhou_2023_CVPR, zhang2024large, bian2024motioncraft, luo2024m3gpt} enabling multiple input modalities.
However, most existing methods focus solely on generative tasks without supporting estimation. 
For instance, the method~\cite{zhang2024large} supports video input but treats it as a generative task, resulting in motions that loosely imitate video content rather than precisely matching it. 
In contrast, our method jointly handles generation and estimation tasks, yielding more precise video-conditioned results.

For long-sequence motion generation, existing works mostly rely on ad-hoc post-processing techniques to stitch separately generated fixed-length motions~\cite{athanasiou22teach,zhange2023diffcollage,Qian_2023_ICCV,petrovich24stmc}. In contrast, our method introduces a novel diffusion-based architecture enabling seamless generation of arbitrary-length motions conditioned on multiple modalities without complex post-processing.

Existing datasets, such as AMASS~\cite{AMASS:ICCV:2019}, are limited in size and diversity. To address the scarcity of 3D data, Motion-X~\cite{lin2023motionx} and MotionBank~\cite{xu2024motionbank} augment datasets using 2D videos and 3D pose estimation models~\cite{yuan2022glamr,shin2024wham}, but the resulting motions often contain artifacts.  In contrast, our method directly leverages in-the-wild videos with 2D annotations without explicit 3D reconstruction, reducing reliance on noisy data and enhancing robustness and diversity.

\subsection{Human Motion Estimation}
Human pose estimation from images~\cite{hmrKanazawa18, li2020hybrik, sarandi2024nlf}, videos~\cite{kocabas2020vibe, choi2020beyond, goel2023humans}, or even sparse marker data~\cite{lee2024mocapevery, TransPoseSIGGRAPH2021, ponton2023sparseposer} has been studied extensively in the literature. Recent works focus primarily on estimating global human motion in world-space coordinates~\cite{yuan2022glamr, ye2023slahmr, kocabas2024pace, li2024coin, shin2024wham, wang2024tram}. This is an inherently ill-posed problem, hence these methods leverage generative priors and SLAM methods to constrain human and camera motions, respectively. 
However, these methods typically involve computationally expensive optimization or separate post-processing steps.

More recent approaches aim to estimate global human motion in a feed-forward manner~\cite{shin2024wham, zhang2024rohm, wang2024tram, shen2024world}, offering faster solutions. Our method extends this direction by jointly modeling generation and estimation within a unified diffusion framework. This integration leverages shared representations and generative priors during training to produce more plausible estimations. 

\section{Generalist Model for Human Motion}

GENMO unifies motion estimation and generation by formulating both tasks as conditional motion generation. Specifically, it synthesizes a human motion sequence $x$ of length $N$ based on a set of condition signals $\mathcal{C}$ and a set of corresponding condition masks $\mathcal{M}$, where $N$ can be arbitrarily large. The condition set $\mathcal{C}$ includes one or more of the following: video feature $c_\text{video}\in \mathbb{R}^{N \times d_\text{video}}$, camera motion $c_\text{cam}\in \mathbb{R}^{N \times d_\text{cam}}$, 2D skeleton $c_\text{2d}\in \mathbb{R}^{N \times d_\text{2d}}$, music clip $c_\text{music}\in \mathbb{R}^{N\times d_\text{music}}$, 2d bounding box $c_\text{bbox}\in \mathbb{R}^{N \times d_\text{bbox}}$, or natural language $c_\text{text}\in \mathbb{R}^{M \times d_\text{text}}$ that describes the motion where $M$ is the number of text tokens. The condition mask  $\mathcal{M}$ consists of the mask $m_\star \in \mathbb{R}^{N \times d_\star}$ for each condition type $c_\star$ in $\mathcal{C}$. The mask matrix is of the same size as the condition feature and its element is one if the condition feature is available and zero otherwise.

\vspace{2mm}
\noindent\textbf{Joint Local and Global Motion Representation.}
We now introduce the motion representation we use for $x$. Most text-to-motion generation methods adopt an egocentric motion representation that encodes human motion in a heading-free local coordinate system. However, for motion estimation, human motions are typically represented in the camera coordinate system to ensure better image feature alignment that facilitates learning. In this work, to obtain a unified generation and estimation model, we adopt a general human motion representation that encodes both the egocentric and camera-space human motions, along with the camera poses.
Our approach leverages the gravity-view coordinate system~\cite{shen2024world}, where the global trajectory of a person at frame $i$ includes the gravity-view orientation $\Gamma^i_\text{gv} \in \mathbb{R}^{6}$ and the local root velocities $v^i_\text{root} \in \mathbb{R}^{3}$. The local motion at the $i$-th frame is represented as the SMPL~\cite{loper2015smpl} parameters, which consists of joint angles $\theta^i \in \mathbb{R}^{24 \times 6}$, shape parameters $\beta^i \in \mathbb{R}^{10}$, and root translation $t^i_\text{root} \in \mathbb{R}^{3}$. Camera pose information at frame $i$ is encoded through the camera-to-world transformation $\pi^i = \big(\Gamma^i_\text{cv}, t^i_\text{cv}\big)$, comprising the camera-view orientation $\Gamma^i_\text{cv} \in \mathbb{R}^{6}$ and camera translation $t^i_\text{cv} \in \mathbb{R}^{3}$. Additionally, we include contact labels $p^i \in \mathbb{R}^{6}$ for hands and feet (heels and toes). The complete motion sequence $x = \big\{{x^i}\big\}_{i=1}^N$ encompasses $N$ human poses, where each pose $x^i \in \mathbb{R}^{D}$ consists of global motion, local motion, and camera pose:
\begin{equation}
    x^i = \big(\Gamma^i_\text{gv}, v^i_\text{root}, \theta^i, \beta^i, t^i_\text{root}, \pi^i, p^i\big).
\end{equation}

\subsection{Unified Estimation and Generation Design}
\begin{figure}[t]
    \centering
    \includegraphics[width=\linewidth]{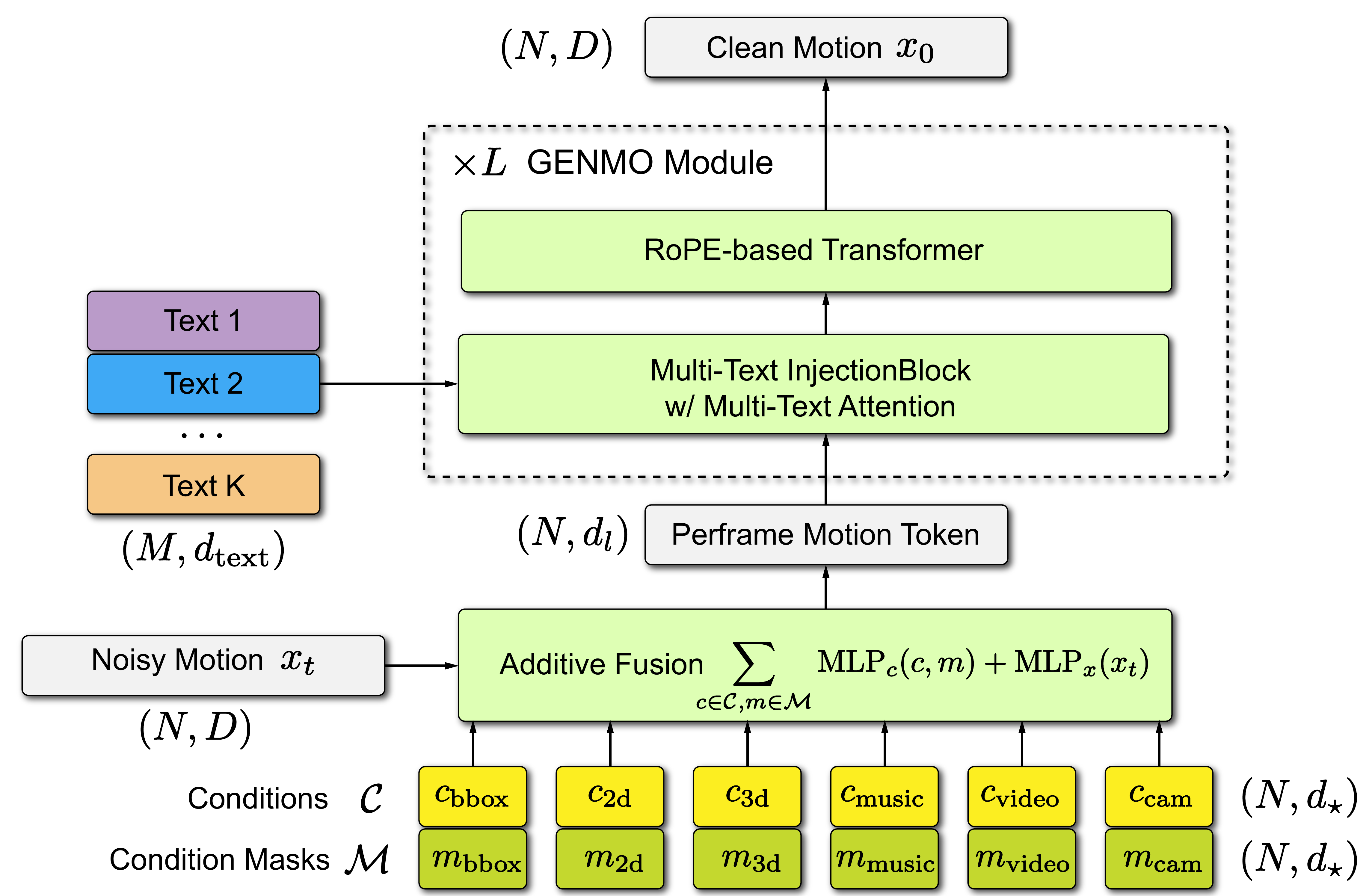}
    \vspace{-5mm}
    \caption{
        \textbf{GENMO Model Design} supports the generation of variable-length motion sequences in a single pass and enables seamless integration of multimodal conditioning signals,  supporting both human motion generation and estimation. 
    }
    \vspace{-2mm}
    \label{fig:overview}
\end{figure}

In this section, we will present the architectural design of GENMO and elucidates how it unifies motion estimation and generation within a single model. The model architecture, illustrated in Figure~\ref{fig:overview}, transforms a noisy motion sequence $x_t$ with the conditions $\mathcal{C}$ and condtion masks $\mathcal{M}$ into a clean motion sequence $x_0$ through a series of carefully designed components. The initial processing stage consists of an additive fusion block that converts $x_t$ into a sequence of per-frame motion tokens. This block utilizes dedicated multilayer perceptrons (MLPs) to process each condition type in $\mathcal{C}$ independently, combines their features through summation to create a unified condition representation, which is further fused with noisy motion $x_t$ to produce the motion token sequence. The resulting sequence is subsequently processed through $L$ GENMO modules, each comprising a RoPE-based Transformer block and our novel multi-text injection block. Our architecture leverages Rotary Position Embedding (RoPE)~\cite{su2024roformer}, which computes attention based on relative temporal positions. This design choice enables processing of variable-length sequences and accommodates conditions lacking inherent temporal ordering, such as images and 2D skeletons.

\begin{figure}[t]
    \centering
    \includegraphics[width=\linewidth]{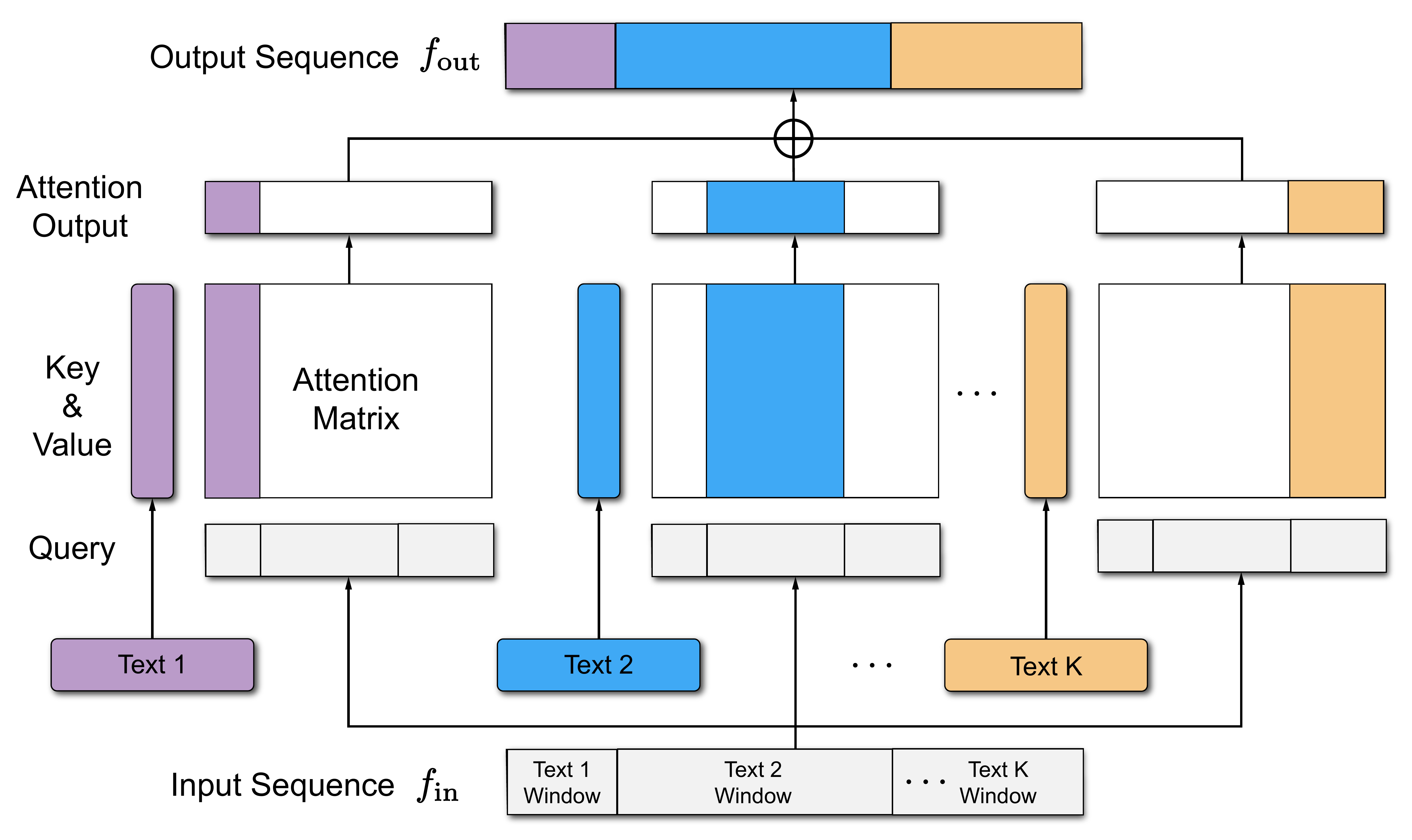}
    \vspace{-7mm}
    \caption{
        \textbf{Multi-text attention} enables flexible conditioning with multiple text inputs, each constrained to its specified time window.
    }
    \vspace{-1mm}
    \label{fig:attention}
\end{figure}

However, text conditioning poses unique challenges. Unlike frame-aligned modalities such as video and music, text is not aligned with the motion frames. The conventional approach of concatenating text with the motion sequence is inadequate as inserting text at any positions can introduce temporal bias.
To address this challenge, we propose a novel multi-text injection block that facilitates text-conditioned motion generation while accommodating multiple text inputs ($K$) with user-specified time windows. The multi-text injection block comprises a transformer block with our proposed multi-text attention mechanism at its core. As depicted in Figure~\ref{fig:attention}, the multi-text attention mechanism processes K text embedding sequences ${c_\text{text}^1, c_\text{text}^2, \ldots, c_\text{text}^K}$ alongside the input motion feature sequence $f_\text{in}$ to generate the output feature sequence $f_\text{out}$:
\begin{align}
    f_{out} & = \sum_{k=1}^K \text{MaskedMHA}\big( f_{in},c_\text{text}^k,\Omega_k\big). \\
    \Omega_k(i,j) & = 
    \begin{cases}
        1 & \text{if }  i \text{ is within time window of text } k \\
        0 & \text{otherwise}
    \end{cases}
\end{align}
where $\text{MaskedMHA}(\cdot)$ represents a masked variant of the conventional multi-head attention mechanism. For each text input $k$, we employ a binary mask $\Omega_k$ that assumes a value of one when timestep $i$ lies within the designated time window of text $k$, and zero otherwise. Through the multiplication of attention weights with mask $\Omega_k$, we effectively restrict the influence of each text prompt to its corresponding time window. Although the mask introduces discontinuities at time window boundaries, GENMO successfully generates smooth motion sequences through the subsequent RoPE-based transformer block, which effectively captures and models temporal motion dynamics.

\vspace{2mm}
\noindent\textbf{Inference with Arbitrary Motion Length.} Our architecture employs relative positional embeddings rather than absolute embeddings for motion sequences, allowing GENMO to generate motions of arbitrary length in a single diffusion forward pass while naturally incorporating multiple text inputs across different time spans. During inference, we adopt a sliding window attention mechanism in the RoPE-based Transformer block, where each token attends only to tokens within a $W$-frame neighborhood. This design enables the generation of motion sequences longer than those seen during training while preserving computational efficiency and ensuring smooth, coherent motion transitions.

\vspace{2mm}
\noindent\textbf{Mixed Multimodal Conditions.} When conditioned on multiple modalities, our framework employs a principled approach for generation: text conditions, which lack frame-level alignment, are processed through our specialized multi-text attention mechanism, while frame-aligned modalities (e.g., video, music, 2D skeleton) are managed through a temporal masking strategy. As mentioned before, for each condition $c_\star$, we use a mask $m_\star$ of the same size to indicate whether the condition feature is (partially) present at each frame (one for present, zero otherwise). 
We also multiply the mask with the condition feature  to nullify missing features.
This simple yet effective approach enables seamless transitions between different conditioning modalities while maintaining temporal coherence.

\subsection{Dual-Mode Training Paradigm}

\begin{figure}[t]
    \centering
    \includegraphics[width=\linewidth]{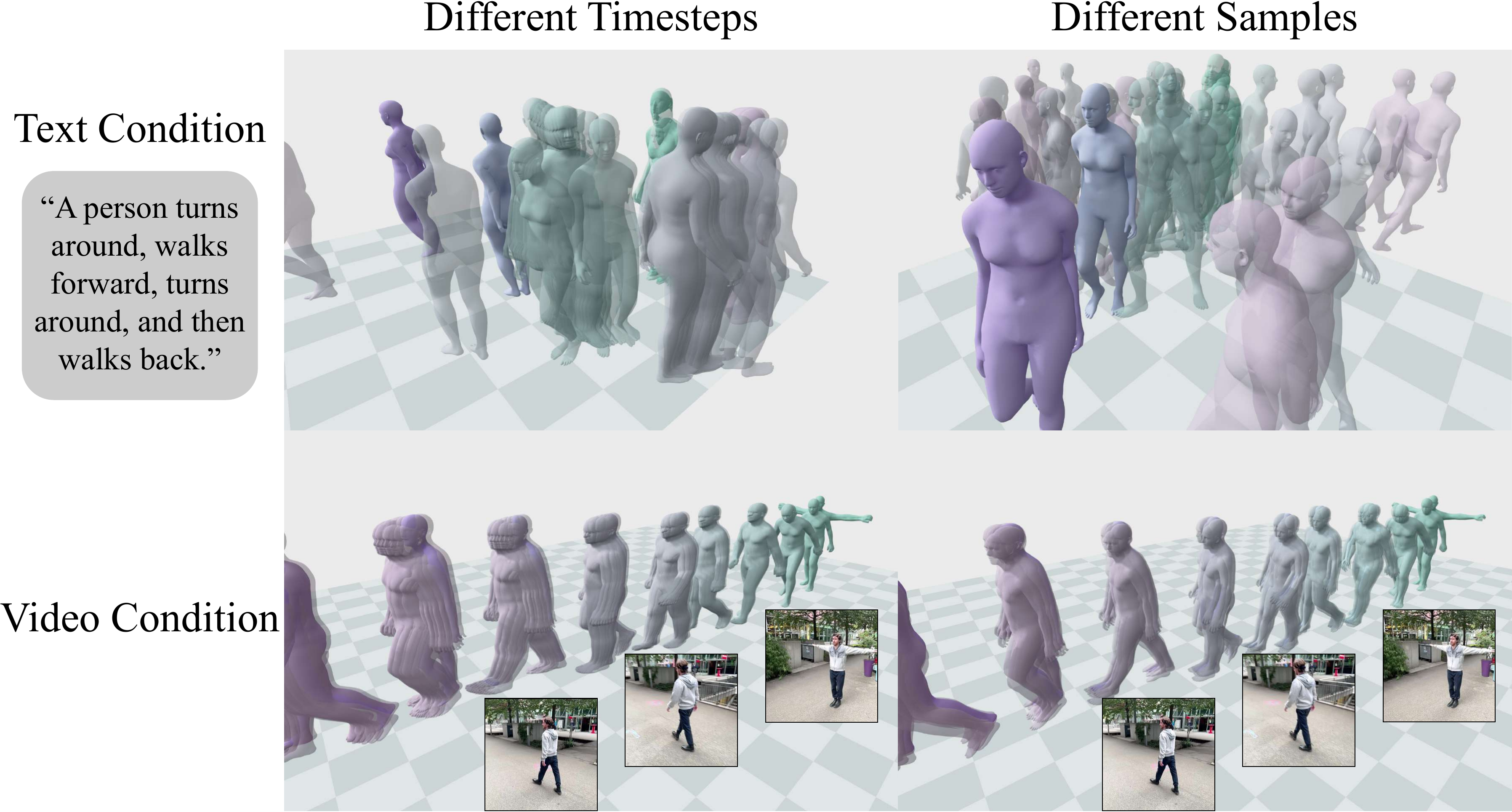}
    \caption{
        \textbf{Variance of video/text conditioned predictions.} Left: Intermediate predictions across 50 DDIM denoising steps. Right: Predictions with 10 different initial noises (including zero-noise). Motions are transparent except the first-step and zero-noise predictions. Video conditioning yields more deterministic outputs compared to text conditioning. \vspace{-5mm}
    }
    \label{fig:var}
\end{figure}

As a diffusion model, GENMO can theoretically be trained with the standard DDPM~\cite{ho2020denoising} objective:
\begin{equation}
    \mathcal{L}_\text{gen} = \mathbb{E}_{t \sim [1, T], x_t \sim q(x_t | x_0)} \big[ \big\| x_0 - \mathcal{G}(x_t, t, \mathcal{C}, \mathcal{M}) \big\|^2 \big]\,,
    \label{eq:ddpm}
\end{equation}
where $t$ the sampled diffusion timestep, and $x_t$ is the noisy motion sampled from the forward diffusion process. Ideally, the model trained with this objective should be capable of generating motion sequences that satisfy the condition set $\mathcal{C}$ and mask $\mathcal{M}$, so it can be used as a motion estimation model when provided with video $c_\text{video}$ or 2D skeleton $c_\text{2d}$ conditions. However, we found that such a generative training objective is not enough to generate accurate motion sequences that are consistent with the input video. We observe a fundamental difference between motion estimation and text-to-motion generation tasks: motion estimation results exhibit substantially lower variability. To investigate this phenomenon, we trained separate diffusion models for text-to-motion generation and video-conditioned motion estimation, then visualized their predictions across all diffusion steps and different initial latent noises (Fig.~\ref{fig:var}). The results demonstrate that the video-conditioned model behaves more deterministically, in other words, the first-step prediction closely resembles predictions from subsequent steps with minimal variation. In contrast, the text-to-motion model exhibits significantly higher variance among steps. This observation has important implications for the estimation task: the accuracy of the first-step prediction becomes critical, as errors introduced early in the diffusion process are difficult to correct in later steps. Based on this insight, we propose a \textit{dual-mode} training paradigm, which consists of (1) an \textit{estimation mode} and (2) a \textit{generation mode}. Intuitively, this dual-mode approach reinforces the quality of first-step predictions while maintaining the model's generative capabilities.

\vspace{2mm}
\noindent\textbf{Estimation Mode.}
In the estimation mode, we formulate the problem as a regression task, employing maximum likelihood estimation to learn the conditional distribution $q(x|\mathcal{C}, \mathcal{M})$. This approach yields the following mean-square error (MSE) objective:
\begin{equation}
    \mathcal{L}_\text{est} = \mathbb{E}_{z \sim \mathcal{N}(\mathbf{0}, {I})} \big[ \big\| x_0 - \mathcal{G}(z, T, \mathcal{C}, \mathcal{M}) \big\|^2 \big] \,.
\end{equation}
Rather than using noisy motion $x_t$, we utilize pure Gaussian noise $z \sim \mathcal{N}(\mathbf{0}, {I})$ as input to the model, along with the corresponding maximum diffusion timestep $T$. This formulation ensures that the estimation mode aligns with the inherent variance characteristics of the diffusion model, thereby preventing conflicts between the generation and estimation modes.

To further enhance the quality of predicted motion sequences, we incorporate geometric regularization losses $\mathcal{L}_\text{geo}$ following established approaches in the literature~\cite{tevet2023human,rempe2021humor}. It involves decoding the predicted motion sequences into SMPL joints and vertices, followed by the application of constraints on world-space and camera-space vertices positions, world-space and camera-space joint positions, and joint contacts. In scenarios where only 2D annotations are available, we employ a 2D reprojection loss to effectively regularize the predicted motion sequences.

\vspace{2mm}
\noindent\textbf{Generation Mode.}
For data with clean 3D annotations $x_0$, we can directly employ the standard diffusion objective in Eq.~\ref{eq:ddpm} to train the generation mode. In this section, we primarily focus on the more interesting scenario where only 2D annotations are available for the generation mode.

Unlike 3D annotations, 2D pose labels are more readily accessible through manual annotation or by applying robust 2D pose estimators on large-scale video datasets. 2D data also offers greater diversity compared to existing 3D motion capture data, which is constrained by the limited variety of subjects, motions, appearances, and environments.

Due to its inherent estimation capability, GENMO can naturally leverage 2D data for training the generation mode. Specifically, we propose an estimation-guided generation training strategy. First, we generate a pseudo-clean motion from the estimation mode using video or 2D skeleton as conditions: $\hat{x}_0 = \mathcal{G}(z, T, \mathcal{C})$. Subsequently, we sample a noisy motion sequence $\hat{x}_t$ through the forward diffusion process: $q\big(\hat{x}_t|\hat{x}_{0}\big)$.
We then apply a 2D reprojection loss on the predicted clean motion using the 2D keypoint annotations $x_\text{2d}$:
\begin{equation}
    \mathcal{L}_\text{gen-2D} = \mathbb{E}_{\hat{x}_t \sim q(\hat{x}_t | \hat{x}_0), t \sim [1, T]} \big[ \big\| x_\text{2d} - \Pi(\mathcal{G}(\hat{x}_t, t, \mathcal{C})) \big\|^2 \big],
    \label{eq:gen_2d}
\end{equation}
where $\Pi$ represents the 2D projection function. For the generation mode, we also apply the aforementioned geometric losses $\mathcal{L}_\text{geo}$ to regularize the predicted motion sequences.

\vspace{2mm}
\noindent\textbf{Training Mode Selection.}
We train the model on diverse datasets with various types of modalities. When training on datasets with strong conditioning signals that render the motion distribution more deterministic, such as video or 2D skeletons, we utilize both the estimation and generation modes to train GENMO. Conversely, when training on datasets with abstract conditions that result in more generative motion distributions, such as text and music, we exclusively employ the generation mode. This mode selection principle is applied to both 3D and 2D data.

\begin{table*}[t]
    \centering
    \caption{
    \textbf{World-grounded human motion estimation.} We evaluate the global motion quality on the EMDB-2~\cite{kaufmann2023emdb} dataset and RICH~\cite{huang2022cap}.
    Parenthesis denotes the number of joints used to compute WA-MPJPE$_{100}$, W-MPJPE$_{100}$ and Jitter.
    }
    \vspace{-3mm}
    \label{tab:quant_global_est}
    \setlength{\tabcolsep}{3pt}
    \resizebox{0.8\textwidth}{!}
    {
    {
            \begin{tabular}{l|ccccc|ccccc}
                \cmidrule[0.75pt]{1-11}
                                                   & \multicolumn{5}{c}{EMDB (24)} & \multicolumn{5}{c}{RICH (24)}                                                                                                                                                                                                          \\
                \cmidrule(lr){2-6} \cmidrule(lr){7-11}

                Models                             & \small{WA-MPJPE$_{100}$} & \small{W-MPJPE$_{100}$}  & \small{RTE} & \small{Jitter} & \small{Foot-Sliding} & \small{WA-MPJPE$_{100}$} & \small{W-MPJPE$_{100}$} & \small{RTE} & \small{Jitter} & \small{Foot-Sliding} \\
                \cmidrule{1-11}
                GLAMR~\cite{yuan2022glamr}                 		& 280.8                         & 726.6                        & 11.4             & 46.3                & 20.7                    & 129.4                         & 236.2                         & 3.8            & 49.7              & 18.1   \\
                TRACE~\cite{sun2023trace}                    		& 529.0                         & 1702.3                       & 17.7             & 2987.6              & 370.7               & 238.1                         & 925.4                         & 610.4        & 1578.6          & 230.7      \\
                SLAHMR~\cite{ye2023slahmr}              	    	& 326.9                         & 776.1                        & 10.2             & 31.3                & 14.5                   & 98.1                           & 186.4                         & 28.9          & 34.3              & 5.1    \\
                COIN~\cite{li2024coin} 					& 152.8 & 407.3 & 3.5 & - & - & - & - & - & - & -\\
                WHAM (w/ DPVO)~\cite{shin2024wham}                & 135.6                         & 354.8                        & 6.0            & 22.5                & 4.4                       & 109.9                         & 184.6                         & 4.1            & 19.7              & 3.3           \\
                WHAM (w/ GT extrinsics)~\cite{shin2024wham}      	& 131.1                         & 335.3                        & 4.1            & 21.0                & 4.4                     & 109.9                         & 184.6                         & 4.1            & 19.7              & 3.3                        \\
                GVHMR (w/ DPVO)~\cite{shen2024world}     	& 111.0                         & 276.5                        & 2.0              & 16.7                & \textbf{3.5}                     & 78.8                           & 126.3                         & 2.4            & \textbf{12.8}              & \textbf{3.0}                      \\
                GVHMR (w/ GT extrinsics)~\cite{shen2024world}      	& 109.1                         & 274.9                        & 1.9              & \textbf{16.5}                & \textbf{3.5}                     & 78.8                           & 126.3                         & 2.4            & \textbf{12.8}              & \textbf{3.0}                       \\
                TRAM (w/ DROID-SLAM)~\cite{wang2024tram} 	& 76.4 & 222.4 & 1.4 & - & - & - & - & - & - & - \\
                \cmidrule{1-11}
                Ours (w/ DROID-SLAM) 				& {74.3} & {202.1} & {1.2} & {17.8} & {8.8} & \textbf{75.3} & \textbf{118.6} & \textbf{1.9} & 15.0 & 6.7  \\
                Ours (w/ GT extrinsics) 					& \textbf{69.5} & \textbf{185.9} & \textbf{0.9} & 17.7 & 8.6 & \textbf{75.3} & \textbf{118.6} & \textbf{1.9} & 15.0 & 6.7  \\
                \cmidrule[0.75pt]{1-11}
            \end{tabular}
            }}
            \vspace{-3mm}
\end{table*}

\begin{table*}[t]
    \centering
    \caption{
    \textbf{Camera-space metrics.} We evaluate the camera-space motion quality on the 3DPW~\cite{vonMarcard2018}, RICH~\cite{huang2022cap} and EMDB-1~\cite{kaufmann2023emdb} datasets.
    $^*$ denotes models trained with the 3DPW training set.
    }
    \vspace{-3mm}
    \label{tab:quant_incam}
    \setlength{\tabcolsep}{3pt}
    \resizebox{0.7\textwidth}{!}
    {\small{
            \begin{tabular}{cl|cccc|cccc|cccc}
                \cmidrule[0.75pt]{1-14}
                 &                           & \multicolumn{4}{c}{3DPW (14)} & \multicolumn{4}{c}{RICH (24)} & \multicolumn{4}{c}{EMDB (24)}                                                                                                                                                                                                \\
                \cmidrule(lr){3-6} \cmidrule(lr){7-10} \cmidrule(lr){11-14}

                 & Models                    & \scriptsize{PA-MPJPE}         & \scriptsize{MPJPE}            & \scriptsize{PVE}              & \scriptsize{Accel} & \scriptsize{PA-MPJPE} & \scriptsize{MPJPE} & \scriptsize{PVE} & \scriptsize{Accel} & \scriptsize{PA-MPJPE} & \scriptsize{MPJPE} & \scriptsize{PVE} & \scriptsize{Accel} \\
                \cmidrule{1-14}

                \multirow{4}{1em}{\rotatebox[origin=c]{90}{per-frame}}
                 & CLIFF$^*$~\cite{li2022cliff}    & 43.0                          & 69.0                          & 81.2                          & 22.5               & 56.6                  & 102.6              & 115.0            & 22.4               & 68.1                  & 103.3              & 128.0            & 24.5               \\
                 & HybrIK$^*$~\cite{li2020hybrik}  & 41.8                          & 71.6                          & 82.3                          & --                 & 56.4                  & 96.8               & 110.4            & --                 & 65.6                  & 103.0              & 122.2            & --                 \\
                 & HMR2.0~\cite{goel2023humans}        & 44.4                          & 69.8                          & 82.2                          & 18.1               & 48.1                  & 96.0               & 110.9            & 18.8               & 60.6                  & 98.0               & 120.3            & 19.8               \\
                 & ReFit$^*$~\cite{refit}    & 40.5                          & 65.3                          & 75.1                          & 18.5               & 47.9                  & 80.7               & 92.9             & 17.1               & 58.6                  & 88.0               & 104.5            & 20.7               \\
                \cmidrule{1-14}

                \multirow{10}{1em}{\rotatebox[origin=c]{90}{temporal}}
                 & VIBE$^*$~\cite{kocabas2020vibe}      & 51.9                          & 82.9                          & 98.4                          & 18.5               & 68.4                  & 120.5              & 140.2            & 21.8               & 81.4                  & 125.9              & 146.8            & 26.6               \\
                 & TRACE$^*$~\cite{sun2023trace}    & 50.9                          & 79.1                          & 95.4                          & 28.6               & --                    & --                 & --               & --                 & 70.9                  & 109.9              & 127.4            & 25.5               \\
                 & SLAHMR~\cite{ye2023slahmr}      & 55.9                          & --                            & --                            & --                 & 52.5                  & --                 & --               & 9.4                & 69.5                  & 93.5               & 110.7            & 7.1                \\
                 & PACE~\cite{kocabas2024pace}          & --                            & --                            & --                            & --                 & 49.3                  & --                 & --               & 8.8                & --                    & --                 & --               & --                 \\
                 & WHAM$^*$~\cite{shin2024wham}      & {35.9}                 & 57.8                          & 68.7                          & 6.6                & 44.3                  & 80.0               & 91.2             & 5.3                & 50.4                  & 79.7               & 94.4             & 5.3                \\
                 & GVHMR$^*$~\cite{shen2024world}                  & 36.2                          & {55.6}                 & {67.2}                 & {5.0}       & {39.5}         & \textbf{66.0}      & \textbf{74.4}    & {4.1}       & {42.7}         & \textbf{72.6}      & \textbf{84.2}    & \textbf{3.6}       \\
                 & TRAM$^*$~\cite{wang2024tram} & 35.6 & 59.3 & 69.6 & {4.9} & - & - & - & - & 45.7 & 74.4 & 86.6 & 4.9 \\
                \cmidrule{2-14}
                 & Ours$*$ (w/o 2D Training) & 35.2 & 55.4 & 67.0 & \textbf{4.8} & 40.6 & 66.4 & 75.4 & \textbf{4.0} & 44.3 & 76.0 & 88.9 & 4.3 \\
                 & Ours$^*$ & \textbf{34.6} & \textbf{53.9} & \textbf{65.8} & {5.2} & \textbf{39.1} & {66.8} & {75.4} & {4.1} & \textbf{42.5} & {73.0} & {84.8} & {3.8} \\
                \cmidrule[0.75pt]{1-14}
            \end{tabular}
        }}
        \vspace{-5mm}

\end{table*}

\section{Experiments}

We evaluate the performance of GENMO on four different tasks including video-to-motion, music-to-dance, text-to-motion, and motion in-betweening. Note that for all experiments we use a single one-in-all checkpoint jointly trained for all tasks unless stated otherwise.

\vspace{2mm}
\noindent\textbf{Datasets.} GENMO is trained on a diverse collection of motion datasets spanning multiple tasks: (1) motion capture data from AMASS~\cite{AMASS:ICCV:2019}; (2) motion estimation benchmarks including BEDLAM~\cite{Black_CVPR_2023}, Human3.6M~\cite{h36m_pami}, and 3DPW~\cite{vonMarcard2018}; (3) music-to-dance data from AIST++~\cite{li2021ai}; (4) text-to-motion data from HumanML3D~\cite{Guo_2022_CVPR}; (5) 2D keypoints and text descriptions from Motion-X~\cite{lin2023motionx}.
Comprehensive details regarding the training procedure and implementation are provided in Appendix~\ref{sec:implementation}.

For evaluation, we use RICH~\cite{huang2022cap}, and EMDB~\cite{kaufmann2023emdb} for global human motion estimation, 3DPW~\cite{vonMarcard2018} for local human motion estimation, AIST++~\cite{li2021ai} for music-to-dance generation, and HumanML3D~\cite{Guo_2022_CVPR} and Motion-X~\cite{lin2023motionx} for text-to-motion generation.

\vspace{2mm}
\noindent\textbf{Evaluation Metrics.} 
For music-to-dance generation, we follow the standard evaluation metrics~\cite{li2021ai,tseng2023edge} and report the FID, Diversity, PFC, and BAS. For text-to-motion generation, we follow the standard evaluation metrics in previous works~\cite{tevet2023human,Guo_2022_CVPR} and report the R-Precision (Top 3), FID, Diversity, and MultiModal Dist.
We also test the motion in-betweening performance by reporting the WA-MPJPE and PA-MPJPE for all the keyframes.

For motion estimation, we report MPJPE, PA-MPJPE, and PVE to evaluate the local motion. Acceleration error (Accel) is also reported to measure the smoothness of the motion. For global motion estimation, we report W-MPJPE$_\text{100}$ and WA-MPJPE$_\text{100}$.
We also evaluate the error accumulation over long sequences by reporting RTE in $\%$.
Jitters and foot sliding (FS) during contacts are also reported.

\vspace{2mm}
\noindent\textbf{Qualitative Results.} We provide extensive qualitative results in the \href{https://research.nvidia.com/labs/dair/genmo}{supplementary videos}, demonstrating the effectiveness and versatility of GENMO. 

\subsection{Evaluation of Motion Estimation}

\paragraph{Global Motion Estimation.} We compare GENMO with state-of-the-art (SOTA) methods for recovering global human motion from videos with dynamic cameras. To ensure fair comparison across methods that employ different SLAM techniques during inference, we also report results using ground-truth camera parameters provided by the datasets. As shown in Table~\ref{tab:quant_global_est}, GENMO consistently outperforms specialized methods trained exclusively for human motion estimation. Notably, our approach achieves a W-MPJPE of $202.1$ mm on the EMDB dataset, surpassing TRAM~\cite{wang2024tram} ($222.4$ mm) despite both methods utilizing identical SLAM systems and backbone features for video encoding. This performance advantage stems from our unified motion generation and estimation framework, where the generative prior enhances the quality of reconstructed motions. GENMO also demonstrates superior performance on the RICH dataset compared to all existing methods. Extensive qualitative results are provided in the \href{https://research.nvidia.com/labs/dair/genmo}{supplementary videos}.

\begin{table}[t]
    \centering
    \caption{
        \textbf{Benchmark of Music-to-Dance Generation.} Motion quality is evaluated on the AIST++~\cite{li2021ai} dataset. 
    }
        \vspace{-3mm}

    \label{tab:quant_music}
    \setlength{\tabcolsep}{10pt} %
    \renewcommand{\arraystretch}{1.2} %
    \resizebox{\columnwidth}{!}{ %
        \begin{tabular}{l|cccccc}
            \toprule
            \textbf{Methods} & \textbf{FID$_k$~$\downarrow$} & \textbf{FID$_m$~$\downarrow$} & \textbf{Div$_k$~$\uparrow$} & \textbf{Div$_m$~$\uparrow$} & \textbf{PFC}~$\downarrow$ & \textbf{BAS}~$\uparrow$ \\
            \midrule
            FACT~\cite{li2021ai} & 86.43 & 43.46 & 6.85 & 3.32 & 2.2543 & 0.1607  \\
            Bailando~\cite{siyao2022bailando} & 28.16 & \textbf{9.62} & 7.83 & 6.34 & 1.754 & 0.2332 \\
            EDGE~\cite{tseng2023edge} & 42.16 & 22.12 & 3.96 & 4.61 & 1.5363 & 0.2334 \\
            \midrule
            Ours (music only) & \textbf{16.10} & 13.91 & 8.47 & 7.26 & 0.7340 & 0.2282 \\
            Ours & 40.91 & 18.51 & \textbf{10.09} & \textbf{7.48} & \textbf{0.3702} & \textbf{0.2708} \\
            \bottomrule
        \end{tabular}
    }
        \vspace{-2mm}
\end{table}

\vspace{2mm}
\noindent\textbf{Local Motion Estimation.} We evaluate GENMO against SOTA methods for local 3D human motion estimation. Quantitative results in Table~\ref{tab:quant_incam} demonstrate that GENMO surpasses existing approaches across most metrics. Additionally, we present results without training on 2D-only data, where the observed performance degradation highlights the effectiveness of our estimation-guided 2D training objective. Further evaluation on the challenging 3DPW-XOCC dataset~\cite{li2023niki} reveals that our generative prior enables GENMO to maintain robust performance even under severe occlusions and truncations. Comprehensive analyses and results on 3DPW-XOCC are provided in Appendix~\ref{sec:eval_3dpw_xocc}.

\subsection{Evaluation of Motion Generation}

\noindent\textbf{Comparison on Music-to-Dance.} We evaluate music-to-dance generation performance on the AIST++ dataset~\cite{li2021ai}, with results presented in Table~\ref{tab:quant_music}. GENMO is benchmarked against SOTA methods and a specialized variant of our model trained exclusively on AIST++ for music-to-dance generation. Notably, our generalist model, jointly trained across multiple estimation and generation tasks, demonstrates substantially enhanced motion diversity, physical plausibility, and motion-music correlation, as evidenced by superior Div$_k$, Div$_m$, PFC, and BAS metrics. While GENMO exhibits higher FID values compared to the specialized music-only variant, this performance differential is expected given that our generalist model was trained on considerably more heterogeneous motion data spanning multiple tasks and domains.

\begin{table}[t]
    \centering
    \caption{
        \textbf{Benchmark of Text-to-Motion Generation} on the HumanML3D~\cite{Guo_2022_CVPR} dataset. R@3 denotes R-Precision (Top 3).
    }
    \vspace{-3mm}
    \label{tab:quant_gen}
    \setlength{\tabcolsep}{10pt} %
    \renewcommand{\arraystretch}{1.2} %
    \resizebox{\columnwidth}{!}{ %
        \begin{tabular}{l|c|cccc}
            \toprule
            \textbf{Methods} & Rep. & \textbf{R@3~$\uparrow$} & \textbf{FID~$\downarrow$} & \textbf{MM Dist~$\downarrow$} & \textbf{Diversity~$\rightarrow$} \\
            \midrule
            Real & HumanML3D & 0.797 & 0.002 & 2.974 & 9.503 \\
            \midrule
            T2M~\cite{guo2022generating} & HumanML3D & 0.740 & 1.067 & 3.340 & 9.188 \\
            MDM~\cite{tevet2023human}   & HumanML3D & 0.611 & 0.544 & 5.566 & 9.559 \\
            M2DM~\cite{kong2023priority} & HumanML3D & 0.763 & 0.352 & 3.134 & 9.926 \\
            EMDM~\cite{zhou2024emdm} & HumanML3D & 0.786 & 0.112 & 3.110 & 9.551 \\ 
            \midrule
            Ours (w/o 2D Training)  & SMPL & 0.556 & 0.245 & 3.128 & 11.660 \\
            Ours  & SMPL & 0.632 & 0.216 & 3.466 & 11.342 \\
            \bottomrule
        \end{tabular}
    }
            \vspace{-3mm}
\end{table}

\begin{table}[t]
    \centering
    \caption{
        \textbf{Benchmark of Text-to-Motion Generation.} Motion quality is evaluated on the Motion-X~\cite{lin2023motionx} dataset. 
    }
        \vspace{-3mm}

    \label{tab:quant_gen_motionx}
    \setlength{\tabcolsep}{10pt} %
    \renewcommand{\arraystretch}{1.2} %
    \resizebox{\columnwidth}{!}{ %
        \begin{tabular}{l|cccc}
            \toprule
            \textbf{Methods} & \textbf{R@3~$\uparrow$} & \textbf{FID~$\downarrow$} & \textbf{MM Dist~$\downarrow$} & \textbf{Diversity~$\rightarrow$} \\
            \midrule
            Real & 0.791 & 0.001 & 2.823 & 11.702 \\
            \midrule
            MDM~\cite{tevet2023human}   & 0.313 & 2.389 & 6.745 & 8.720 \\
            Ours (w/o 2D Training) & 0.401 & 0.515 & 5.210 & 12.124 \\
            Ours  &  \textbf{0.472} & \textbf{0.207} & \textbf{4.801} & 11.719   \\      \bottomrule
        \end{tabular}
    }
        \vspace{-3mm}
\end{table}

\vspace{2mm}
\noindent\textbf{Comparison on Text-to-Motion.} We evaluate the text-to-motion generation capabilities of GENMO on both HumanML3D (Table~\ref{tab:quant_gen}) and Motion-X (Table~\ref{tab:quant_gen_motionx}) datasets. Our method demonstrates superior performance compared to the baseline model MDM~\cite{tevet2023human}, exhibiting enhanced motion fidelity and improved text-prompt correspondence across both benchmarks.
To assess the impact of 2D data training, we compare GENMO with its variant without training on Motion-X's 2D data. The results indicate that incorporating 2D training substantially enhances motion generation performance across both HumanML3D and Motion-X datasets. These findings substantiate the efficacy of leveraging 2D data within GENMO's framework for text-conditioned motion generation tasks.

\vspace{2mm}
\noindent\textbf{Discussion on HumanML3D Performance.} Although GENMO exhibits worse performance compared to SOTA methods like EMDM~\cite{zhou2024emdm}, this discrepancy can be stemmed from our representation choice: GENMO utilizes SMPL parameters to represent human motion for unified estimation and generation, whereas SOTA methods employ the HumanML3D representation — the same representation used by the encoders of the FID and R-Precision metrics. \textit{This representational mismatch introduces an inherent disadvantage for GENMO}, as it necessitates bidirectional conversion of ground-truth motions from HumanML3D to SMPL during training and conversion of our generated motions to the HumanML3D format during evaluation. These conversion processes inevitably introduce distribution shifts through alterations in bone lengths, joint angles, and joint velocities, consequently affecting performance metrics and limiting the upper bound of GENMO's achievable performance on these HumanML3D-specific metrics.

\begin{table}[t]
    \centering
    \caption{
    \textbf{Motion In-betweening Experiments.} The DDPM baseline is the proposed method without the estimation objective, only using the standard diffusion objective for training. ``w/o Estimation.'' is the proposed method without training for the motion estimation task. ``w/o 2D Training'' is trained without $\mathcal{L}_{\text{gen-2D}}$. Results are reported using PA-MPJPE/WA-MPJPE.
    }
    \vspace{-3mm}
    \label{tab:quant_abla_ibt}
    \setlength{\tabcolsep}{3pt}
    \resizebox{0.85\linewidth}{!}
    {
    {
            \begin{tabular}{l|cc|cc}
                \cmidrule[0.75pt]{1-5}
                ~ & \multicolumn{2}{c}{HumanML3D} & \multicolumn{2}{c}{Motion-X}          \\
                \cmidrule(lr){2-3} \cmidrule(lr){4-5}

                Models & \small{2-Keyframe} & \small{5-Keyframe} & \small{2-Keyframe} & \small{5-Keyframe} \\
                \cmidrule{1-5}
                Diffusion-only & 71.6/98.8 & 46.3/70.4 & 97.6/154.9 & 56.3/106.9 \\
                w/o Estimation & 64.9/97.5 & 47.5/72.6 & 97.9/151.0 & 69.6/116.4 \\
                w/o 2D Training & 56.4/\textbf{85.1} & \textbf{36.7}/59.5 & 68.3/136.8 & 44.6/98.6 \\
                \cmidrule{1-5}
                Ours & \textbf{53.5}/85.3 & 37.1/\textbf{58.5} & \textbf{58.8/122.7} & \textbf{40.5/89.5} \\
                \cmidrule[0.75pt]{1-5}
            \end{tabular}
        }}
        \vspace{-3mm}

\end{table}

\vspace{2mm}
\noindent\textbf{Experiments on Motion In-betweening.} 
We further evaluate the performance of conditional motion generation through the motion in-betweening task, following the methodology of prior diffusion-based approaches~\cite{tevet2023human} by overwriting the noisy motion with the keyframe poses before each denoising step. 
Experiments are conducted on both HumanML3D and Motion-X test sets under two configurations with either 2 or 5 sampled keyframes.
As shown in Table~\ref{tab:quant_abla_ibt}, GENMO achieves superior performance through its unified estimation and generation training compared to the diffusion-only baseline. Furthermore, the incorporation of additional 2D-only data and joint training with video-conditioned motion estimation substantially enhances motion in-betweening quality.

\begin{table}[t]
    \centering
    \caption{
    \textbf{Ablation studies on motion estimation.} The DDPM baseline is the proposed method without the estimation objective, only using the standard diffusion objective for training. The regression baseline is the proposed method without the generation objective.
    }
    \vspace{-3mm}
    \label{tab:quant_abla_est}
    \setlength{\tabcolsep}{3pt}
    \resizebox{0.9\linewidth}{!}
    {
    {
            \begin{tabular}{l|cc|cc}
                \cmidrule[0.75pt]{1-5}
                ~ & \multicolumn{2}{c}{RICH (24)} & \multicolumn{2}{c}{EMDB (24)}          \\
                \cmidrule(lr){2-3} \cmidrule(lr){4-5}

                Models & \small{WA-MPJPE$_{100}$} & \small{W-MPJPE$_{100}$} & \small{WA-MPJPE$_{100}$} & \small{W-MPJPE$_{100}$} \\
                \cmidrule{1-5}
                Diffusion-only & 88.9 & 143.9 & 128.6 & 307.7 \\
                Regression-only & 87.0 & 141.0 & 121.1 & 300.1 \\
                \cmidrule{1-5}
                Ours & \textbf{81.3} & \textbf{130.6} & \textbf{114.6} & \textbf{281.7} \\
                \cmidrule[0.75pt]{1-5}
            \end{tabular}
        }}

\end{table}

\subsection{Ablation Study}

\noindent\textbf{Impact of the Estimation Mode.} To assess the efficacy of our proposed estimation mode, we evaluate a variant of our method trained exclusively with the generation mode (``Diffusion-only''). Table~\ref{tab:quant_abla_est} presents quantitative comparisons of global human motion estimation performance on the RICH and EMDB datasets, using direct model predictions without post-processing for static joints. The results demonstrate that omitting the estimation objective significantly degrades global motion estimation performance, confirming the estimation objective's crucial role in enhancing consistency between predicted motions and input videos. This finding is further corroborated by the motion in-betweening results in Table~\ref{tab:quant_abla_ibt}, which similarly indicate that the estimation objective improves in-betweening performance.

\vspace{2mm}
\noindent\textbf{Impact of the Generation Mode.} We compare our unified model against a pure regression baseline (trained solely with the estimation mode, akin to SOTA human motion estimation methods) to evaluate the impact of the generation objective. Quantitative comparisons on the RICH and EMDB datasets (Table~\ref{tab:quant_abla_est}) reveal that our unified model consistently outperforms the regression baseline, suggesting that incorporating generative priors enhances motion quality in human motion estimation tasks.

\begin{table}[t]
    \centering
    \caption{
    \textbf{Effect of inference steps} on motion generation and estimation performance.
    }
    \vspace{-3mm}
    \label{tab:quant_abla_timesteps}
    \setlength{\tabcolsep}{3pt}
    \resizebox{0.8\linewidth}{!}
    {
    {
            \begin{tabular}{l|c|cc}
                \cmidrule[0.75pt]{1-4}
                ~ & \multicolumn{1}{c}{HumanML3D (Gen.)} & \multicolumn{2}{c}{EMDB (Est.)}          \\
                \cmidrule(lr){2-3} \cmidrule(lr){3-4}
                Models & \small{FID} &  \small{W-MPJPE$_{100}$} & \small{MPJPE} \\
                \cmidrule{1-4}
                Step=1 (Regression) & 0.260$^{\pm .101}$  & 280.0 & 73.0 \\ %
                Step=2 & 0.242$^{\pm .083}$ & 276.8 & 72.5 \\ %
                Step=5 & 0.231$^{\pm .091}$ & \textbf{274.9} & \textbf{72.2} \\ %
                Step=10 & 0.237$^{\pm .126}$ & 275.8 & 72.3 \\ %
                Step=50 & \textbf{0.216}$^{\pm .119}$ & 278.7 & 72.7 \\ %
                \cmidrule[0.75pt]{1-4}
            \end{tabular}
        }}
        \vspace{-1mm}

\end{table}

\vspace{2mm}
\noindent\textbf{Different Inference Steps.}
We evaluate the impact of denoising steps using the standard DDIM~\cite{song2020denoising} inference pipeline. As shown in Table~\ref{tab:quant_abla_timesteps}, motion estimation performance remains relatively stable across different step counts, while text-to-motion generation shows greater sensitivity. Notably, single-step denoising sufficiently produces video-consistent human motions, with optimal estimation performance achieved at 5 inference steps — a balance that effectively leverages generative priors without introducing excessive variance.

\section{Conclusion}
In this work, we introduced GENMO, a generalist framework for human motion modeling that bridges the gap between motion estimation and generation tasks. We showed that GENMO can effectively leverage shared representations to enable synergistic benefits: generative priors enhance motion estimation robustness under challenging conditions, while diverse video data enriches the generative capabilities. GENMO can produce variable-length motion generation in a single pass and supports flexible control using text, videos, music, 2D keypoints, and 3D keyframes. GENMO achieved state-of-the-art performance on both motion estimation and generation benchmarks, while also reducing reliance on 3D motion capture data. Extensive experiments demonstrated that GENMO is not only capable of handling multiple human motion tasks within a single framework but also achieves superior results compared to task-specific models. 

As with any other work, GENMO has some limitations. Currently, it relies on off-the-shelf SLAM methods to obtain camera parameters for videos. Integrating camera estimation inside GENMO is an interesting future work. Moreover, currently, our model only supports full-body motion. We plan to enable facial expressions and hand articulation in the future.

{
    \small
    \bibliographystyle{ieeenat_fullname}
    \bibliography{main}

\begin{thebibliography}{92}
\providecommand{\natexlab}[1]{#1}
\providecommand{\url}[1]{\texttt{#1}}
\expandafter\ifx\csname urlstyle\endcsname\relax
  \providecommand{\doi}[1]{doi: #1}\else
  \providecommand{\doi}{doi: \begingroup \urlstyle{rm}\Url}\fi

\bibitem[Alexanderson et~al.(2023)Alexanderson, Nagy, Beskow, and Henter]{alexanderson2023listen}
Simon Alexanderson, Rajmund Nagy, Jonas Beskow, and Gustav~Eje Henter.
\newblock Listen, denoise, action! audio-driven motion synthesis with diffusion models.
\newblock \emph{ACM Transactions on Graphics ({TOG})}, 2023.

\bibitem[Athanasiou et~al.(2022)Athanasiou, Petrovich, Black, and Varol]{athanasiou22teach}
Nikos Athanasiou, Mathis Petrovich, Michael~J. Black, and G{\"u}l Varol.
\newblock {TEACH}: Temporal action composition for {3D} humans.
\newblock In \emph{International Conference on {3D} Vision ({3DV})}, 2022.

\bibitem[Barsoum et~al.(2018)Barsoum, Kender, and Liu]{BarsoumCVPRW2018}
Emad Barsoum, John Kender, and Zicheng Liu.
\newblock {HP-GAN}: Probabilistic {3D} human motion prediction via {GAN}.
\newblock In \emph{CVPR Workshops}, 2018.

\bibitem[Bian et~al.(2024)Bian, Zeng, Ju, Liu, Zhang, Liu, and Xu]{bian2024motioncraft}
Yuxuan Bian, Ailing Zeng, Xuan Ju, Xian Liu, Zhaoyang Zhang, Wei Liu, and Qiang Xu.
\newblock Motioncraft: Crafting whole-body motion with plug-and-play multimodal controls.
\newblock \emph{arXiv preprint arXiv:2407.21136}, 2024.

\bibitem[Black et~al.(2023)Black, Patel, Tesch, and Yang]{Black_CVPR_2023}
Michael~J. Black, Priyanka Patel, Joachim Tesch, and Jinlong Yang.
\newblock {BEDLAM}: A synthetic dataset of bodies exhibiting detailed lifelike animated motion.
\newblock In \emph{Proceedings IEEE/CVF Conf.~on Computer Vision and Pattern Recognition (CVPR)}, pages 8726--8737, 2023.

\bibitem[Bogo et~al.(2016)Bogo, Kanazawa, Lassner, Gehler, Romero, and Black]{bogo2016keep}
Federica Bogo, Angjoo Kanazawa, Christoph Lassner, Peter Gehler, Javier Romero, and Michael~J Black.
\newblock Keep it {SMPL}: Automatic estimation of {3D} human pose and shape from a single image.
\newblock In \emph{ECCV}, 2016.

\bibitem[Cervantes et~al.(2022)Cervantes, Sekikawa, Sato, and Shinoda]{cervantes2022implicit}
Pablo Cervantes, Yusuke Sekikawa, Ikuro Sato, and Koichi Shinoda.
\newblock Implicit neural representations for variable length human motion generation.
\newblock In \emph{ECCV}, 2022.

\bibitem[Chen et~al.(2023)Chen, Jiang, Liu, Huang, Fu, Chen, Yu, and Yu]{chen2023mld}
Xin Chen, Biao Jiang, Wen Liu, Zilong Huang, Bin Fu, Tao Chen, Jingyi Yu, and Gang Yu.
\newblock Executing your commands via motion diffusion in latent space.
\newblock In \emph{CVPR}, 2023.

\bibitem[Choi et~al.(2021)Choi, Moon, and Lee]{choi2020beyond}
Hongsuk Choi, Gyeongsik Moon, and Kyoung~Mu Lee.
\newblock Beyond static features for temporally consistent 3d human pose and shape from a video.
\newblock In \emph{CVPR}, 2021.

\bibitem[Chopin et~al.(2021)Chopin, Otberdout, Daoudi, and Bartolo]{chopin21_wgan}
B. Chopin, N. Otberdout, M. Daoudi, and A. Bartolo.
\newblock Human motion prediction using manifold-aware wasserstein gan.
\newblock In \emph{FG}, 2021.

\bibitem[Cohan et~al.(2024)Cohan, Tevet, Reda, Peng, and van~de Panne]{cohan2024flexible}
Setareh Cohan, Guy Tevet, Daniele Reda, Xue~Bin Peng, and Michiel van~de Panne.
\newblock Flexible motion in-betweening with diffusion models.
\newblock \emph{SIGGRAPH}, 2024.

\bibitem[Dai et~al.(2024)Dai, Chen, Wang, Liu, Dai, and Tang]{motionlcm}
Wenxun Dai, Ling-Hao Chen, Jingbo Wang, Jinpeng Liu, Bo Dai, and Yansong Tang.
\newblock Motionlcm: Real-time controllable motion generation via latent consistency model.
\newblock In \emph{ECCV}, 2024.

\bibitem[Fu et~al.(2024)Fu, Yin, Hu, Wang, Ma, Tan, Shen, Lin, and Long]{fu2024geowizard}
Xiao Fu, Wei Yin, Mu Hu, Kaixuan Wang, Yuexin Ma, Ping Tan, Shaojie Shen, Dahua Lin, and Xiaoxiao Long.
\newblock Geowizard: Unleashing the diffusion priors for 3d geometry estimation from a single image.
\newblock In \emph{European Conference on Computer Vision}, pages 241--258. Springer, 2024.

\bibitem[Ghosh et~al.(2021)Ghosh, Cheema, Oguz, Theobalt, and Slusallek]{Ghosh_2021_ICCV}
Anindita Ghosh, Noshaba Cheema, Cennet Oguz, Christian Theobalt, and Philipp Slusallek.
\newblock Synthesis of compositional animations from textual descriptions.
\newblock In \emph{ICCV}, 2021.

\bibitem[Goel et~al.(2023)Goel, Pavlakos, Rajasegaran, Kanazawa, and Malik]{goel2023humans}
Shubham Goel, Georgios Pavlakos, Jathushan Rajasegaran, Angjoo Kanazawa, and Jitendra Malik.
\newblock Humans in 4d: Reconstructing and tracking humans with transformers.
\newblock In \emph{Proceedings of the IEEE/CVF International Conference on Computer Vision}, pages 14783--14794, 2023.

\bibitem[Guo et~al.(2020)Guo, Zuo, Wang, Zou, Sun, Deng, Gong, and Cheng]{chuan2020action2motion}
Chuan Guo, Xinxin Zuo, Sen Wang, Shihao Zou, Qingyao Sun, Annan Deng, Minglun Gong, and Li Cheng.
\newblock {Action2Motion}: Conditioned generation of {3D} human motions.
\newblock In \emph{ACM International Conference on Multimedia ({ACMMM})}, 2020.

\bibitem[Guo et~al.(2022{\natexlab{a}})Guo, Zou, Zuo, Wang, Ji, Li, and Cheng]{Guo_2022_CVPR}
Chuan Guo, Shihao Zou, Xinxin Zuo, Sen Wang, Wei Ji, Xingyu Li, and Li Cheng.
\newblock Generating diverse and natural 3d human motions from text.
\newblock In \emph{Proceedings of the IEEE/CVF Conference on Computer Vision and Pattern Recognition (CVPR)}, pages 5152--5161, 2022{\natexlab{a}}.

\bibitem[Guo et~al.(2022{\natexlab{b}})Guo, Zou, Zuo, Wang, Ji, Li, and Cheng]{guo2022generating}
Chuan Guo, Shihao Zou, Xinxin Zuo, Sen Wang, Wei Ji, Xingyu Li, and Li Cheng.
\newblock Generating diverse and natural 3d human motions from text.
\newblock In \emph{Proceedings of the IEEE/CVF conference on computer vision and pattern recognition}, pages 5152--5161, 2022{\natexlab{b}}.

\bibitem[Guo et~al.(2022{\natexlab{c}})Guo, Zuo, Wang, and Cheng]{chuan2022tm2t}
Chuan Guo, Xinxin Zuo, Sen Wang, and Li Cheng.
\newblock {TM2T}: Stochastic and tokenized modeling for the reciprocal generation of 3d human motions and texts.
\newblock In \emph{ECCV}, 2022{\natexlab{c}}.

\bibitem[Guo et~al.(2023)Guo, Mu, Javed, Wang, and Cheng]{guo2023momask}
Chuan Guo, Yuxuan Mu, Muhammad~Gohar Javed, Sen Wang, and Li Cheng.
\newblock Momask: Generative masked modeling of 3d human motions.
\newblock 2023.

\bibitem[Habibie et~al.(2017)Habibie, Holden, Schwarz, Yearsley, and Komura]{Habibie2017ARV}
Ikhsanul Habibie, Daniel Holden, Jonathan Schwarz, Joe Yearsley, and Taku Komura.
\newblock A recurrent variational autoencoder for human motion synthesis.
\newblock In \emph{BMVC}, 2017.

\bibitem[Hassan et~al.(2021)Hassan, Ceylan, Villegas, Saito, Yang, Zhou, and Black]{hassan2021stochastic}
Mohamed Hassan, Duygu Ceylan, Ruben Villegas, Jun Saito, Jimei Yang, Yi Zhou, and Michael~J Black.
\newblock Stochastic scene-aware motion prediction.
\newblock In \emph{ICCV}, 2021.

\bibitem[He et~al.(2022)He, Saito, Zachary, Rushmeier, and Zhou]{he2022nemf}
Chengan He, Jun Saito, James Zachary, Holly Rushmeier, and Yi Zhou.
\newblock Nemf: Neural motion fields for kinematic animation.
\newblock In \emph{NeurIPS}, 2022.

\bibitem[Henter et~al.(2020)Henter, Alexanderson, and Beskow]{Henter2020}
Gustav~Eje Henter, Simon Alexanderson, and Jonas Beskow.
\newblock {MoGlow}: Probabilistic and controllable motion synthesis using normalising flows.
\newblock \emph{ACM Transactions on Graphics ({TOG})}, 2020.

\bibitem[Heusel et~al.(2017)Heusel, Ramsauer, Unterthiner, Nessler, and Hochreiter]{heusel2017gans}
Martin Heusel, Hubert Ramsauer, Thomas Unterthiner, Bernhard Nessler, and Sepp Hochreiter.
\newblock Gans trained by a two time-scale update rule converge to a local nash equilibrium.
\newblock \emph{Advances in neural information processing systems}, 30, 2017.

\bibitem[Ho et~al.(2020)Ho, Jain, and Abbeel]{ho2020denoising}
Jonathan Ho, Ajay Jain, and Pieter Abbeel.
\newblock Denoising diffusion probabilistic models.
\newblock \emph{Advances in neural information processing systems}, 33:\penalty0 6840--6851, 2020.

\bibitem[Huang et~al.(2022)Huang, Yi, H{\"o}schle, Safroshkin, Alexiadis, Polikovsky, Scharstein, and Black]{huang2022cap}
Chun-Hao~P. Huang, Hongwei Yi, Markus H{\"o}schle, Matvey Safroshkin, Tsvetelina Alexiadis, Senya Polikovsky, Daniel Scharstein, and Michael~J Black.
\newblock Capturing and inferring dense full-body human-scene contact.
\newblock In \emph{CVPR}, 2022.

\bibitem[Ionescu et~al.(2014)Ionescu, Papava, Olaru, and Sminchisescu]{h36m_pami}
Catalin Ionescu, Dragos Papava, Vlad Olaru, and Cristian Sminchisescu.
\newblock {Human3.6M}: Large scale datasets and predictive methods for {3D} human sensing in natural environments.
\newblock \emph{TPAMI}, 36\penalty0 (7):\penalty0 1325--1339, 2014.

\bibitem[Jin et~al.(2023)Jin, Wu, Fan, Sun, Wei, and Yuan]{jin2023act}
Peng Jin, Yang Wu, Yanbo Fan, Zhongqian Sun, Yang Wei, and Li Yuan.
\newblock Act as you wish: Fine-grained control of motion diffusion model with hierarchical semantic graphs.
\newblock In \emph{NeurIPs}, 2023.

\bibitem[Kanazawa et~al.(2018)Kanazawa, Black, Jacobs, and Malik]{hmrKanazawa18}
Angjoo Kanazawa, Michael~J. Black, David~W. Jacobs, and Jitendra Malik.
\newblock End-to-end recovery of human shape and pose.
\newblock In \emph{CVPR}, 2018.

\bibitem[Kaufmann et~al.(2023)Kaufmann, Song, Guo, Shen, Jiang, Tang, Z{\'a}rate, and Hilliges]{kaufmann2023emdb}
Manuel Kaufmann, Jie Song, Chen Guo, Kaiyue Shen, Tianjian Jiang, Chengcheng Tang, Juan~Jos{\'e} Z{\'a}rate, and Otmar Hilliges.
\newblock {EMDB}: The {E}lectromagnetic {D}atabase of {G}lobal 3{D} {H}uman {P}ose and {S}hape in the {W}ild.
\newblock In \emph{ICCV}, 2023.

\bibitem[Ke et~al.(2024)Ke, Obukhov, Huang, Metzger, Caye~Daudt, and Schindler]{ke2023repurposing}
Bingxin Ke, Anton Obukhov, Shengyu Huang, Nando Metzger, Rodrigo Caye~Daudt, and Konrad Schindler.
\newblock Repurposing diffusion-based image generators for monocular depth estimation.
\newblock In \emph{CVPR}, 2024.

\bibitem[Kocabas et~al.(2020)Kocabas, Athanasiou, and Black]{kocabas2020vibe}
Muhammed Kocabas, Nikos Athanasiou, and Michael~J Black.
\newblock {VIBE}: Video inference for human body pose and shape estimation.
\newblock In \emph{CVPR}, 2020.

\bibitem[Kocabas et~al.(2021)Kocabas, Huang, Hilliges, and Black]{pare2021kocabas}
Muhammed Kocabas, Chun-Hao~P. Huang, Otmar Hilliges, and Michael~J. Black.
\newblock {PARE}: Part attention regressor for {3D} human body estimation.
\newblock In \emph{ICCV}, 2021.

\bibitem[Kocabas et~al.(2024)Kocabas, Yuan, Molchanov, Guo, Black, Hilliges, Kautz, and Iqbal]{kocabas2024pace}
Muhammed Kocabas, Ye Yuan, Pavlo Molchanov, Yunrong Guo, Michael~J. Black, Otmar Hilliges, Jan Kautz, and Umar Iqbal.
\newblock {PACE}: Human and motion estimation from in-the-wild videos.
\newblock In \emph{3DV}, 2024.

\bibitem[Kong et~al.(2023)Kong, Gong, Lian, Mi, and Wang]{kong2023priority}
Hanyang Kong, Kehong Gong, Dongze Lian, Michael~Bi Mi, and Xinchao Wang.
\newblock Priority-centric human motion generation in discrete latent space.
\newblock In \emph{Proceedings of the IEEE/CVF International Conference on Computer Vision}, pages 14806--14816, 2023.

\bibitem[Kulkarni et~al.(2023)Kulkarni, Rempe, Genova, Kundu, Johnson, Fouhey, and Guibas]{kulkarni2023nifty}
Nilesh Kulkarni, Davis Rempe, Kyle Genova, Abhijit Kundu, Justin Johnson, David Fouhey, and Leonidas Guibas.
\newblock {NIFTY}: Neural object interaction fields for guided human motion synthesis.
\newblock \emph{arXiv:2307.07511}, 2023.

\bibitem[Lee and Joo(2024)]{lee2024mocapevery}
Jiye Lee and Hanbyul Joo.
\newblock Mocap everyone everywhere: Lightweight motion capture with smartwatches and a head-mounted camera.
\newblock In \emph{Proceedings of the IEEE/CVF Conference on Computer Vision and Pattern Recognition (CVPR)}, 2024.

\bibitem[Li et~al.(2021{\natexlab{a}})Li, Xu, Chen, Bian, Yang, and Lu]{li2020hybrik}
Jiefeng Li, Chao Xu, Zhicun Chen, Siyuan Bian, Lixin Yang, and Cewu Lu.
\newblock Hybrik: A hybrid analytical-neural inverse kinematics solution for 3d human pose and shape estimation.
\newblock In \emph{CVPR}, 2021{\natexlab{a}}.

\bibitem[Li et~al.(2023)Li, Bian, Liu, Tang, Wang, and Lu]{li2023niki}
Jiefeng Li, Siyuan Bian, Qi Liu, Jiasheng Tang, Fan Wang, and Cewu Lu.
\newblock Niki: Neural inverse kinematics with invertible neural networks for 3d human pose and shape estimation.
\newblock In \emph{Proceedings of the IEEE/CVF Conference on Computer Vision and Pattern Recognition}, pages 12933--12942, 2023.

\bibitem[Li et~al.(2024)Li, Yuan, Rempe, Zhang, Molchanov, Lu, Kautz, and Iqbal]{li2024coin}
Jiefeng Li, Ye Yuan, Davis Rempe, Haotian Zhang, Pavlo Molchanov, Cewu Lu, Jan Kautz, and Umar Iqbal.
\newblock Coin: Control-inpainting diffusion prior for human and camera motion estimation.
\newblock In \emph{ECCV}, pages 426--446. Springer, 2024.

\bibitem[Li et~al.(2021{\natexlab{b}})Li, Yang, Ross, and Kanazawa]{li2021ai}
Ruilong Li, Shan Yang, David~A Ross, and Angjoo Kanazawa.
\newblock Ai choreographer: Music conditioned 3d dance generation with aist++.
\newblock In \emph{ICCV}, 2021{\natexlab{b}}.

\bibitem[Li et~al.(2022)Li, Liu, Zhang, Xu, and Yan]{li2022cliff}
Zhihao Li, Jianzhuang Liu, Zhensong Zhang, Songcen Xu, and Youliang Yan.
\newblock Cliff: Carrying location information in full frames into human pose and shape estimation.
\newblock In \emph{ECCV}, 2022.

\bibitem[Lin et~al.(2023)Lin, Zeng, Lu, Cai, Zhang, Wang, and Zhang]{lin2023motionx}
Jing Lin, Ailing Zeng, Shunlin Lu, Yuanhao Cai, Ruimao Zhang, Haoqian Wang, and Lei Zhang.
\newblock Motion-x: A large-scale 3d expressive whole-body human motion dataset.
\newblock In \emph{NeurIPS}, 2023.

\bibitem[Loper et~al.(2015)Loper, Mahmood, Romero, Pons-Moll, and Black]{loper2015smpl}
Matthew Loper, Naureen Mahmood, Javier Romero, Gerard Pons-Moll, and Michael~J Black.
\newblock Smpl: a skinned multi-person linear model.
\newblock \emph{ACM Transactions on Graphics (TOG)}, 34\penalty0 (6):\penalty0 1--16, 2015.

\bibitem[Loshchilov et~al.(2017)Loshchilov, Hutter, et~al.]{loshchilov2017fixing}
Ilya Loshchilov, Frank Hutter, et~al.
\newblock Fixing weight decay regularization in adam.
\newblock \emph{arXiv preprint arXiv:1711.05101}, 5, 2017.

\bibitem[Luo et~al.(2024)Luo, Hou, Li, Chang, Liu, Wang, and Shan]{luo2024m3gpt}
Mingshuang Luo, Ruibing Hou, Zhuo Li, Hong Chang, Zimo Liu, Yaowei Wang, and Shiguang Shan.
\newblock M$^3$gpt: An advanced multimodal, multitask framework for motion comprehension and generation.
\newblock In \emph{NeurIPs}, 2024.

\bibitem[Mahmood et~al.(2019)Mahmood, Ghorbani, Troje, Pons-Moll, and Black]{AMASS:ICCV:2019}
Naureen Mahmood, Nima Ghorbani, Nikolaus~F. Troje, Gerard Pons-Moll, and Michael~J. Black.
\newblock {AMASS}: Archive of motion capture as surface shapes.
\newblock In \emph{ICCV}, 2019.

\bibitem[Martin~Garcia et~al.(2025)Martin~Garcia, Abou~Zeid, Schmidt, de~Geus, Hermans, and Leibe]{martingarcia2024diffusione2eft}
Gonzalo Martin~Garcia, Karim Abou~Zeid, Christian Schmidt, Daan de Geus, Alexander Hermans, and Bastian Leibe.
\newblock Fine-tuning image-conditional diffusion models is easier than you think.
\newblock In \emph{WACV}, 2025.

\bibitem[M{\"u}ller et~al.(2005)M{\"u}ller, R{\"o}der, and Clausen]{muller2005efficient}
Meinard M{\"u}ller, Tido R{\"o}der, and Michael Clausen.
\newblock Efficient content-based retrieval of motion capture data.
\newblock In \emph{ACM SIGGRAPH 2005 Papers}, pages 677--685. 2005.

\bibitem[Onuma et~al.(2008)Onuma, Faloutsos, and Hodgins]{onuma2008fmdistance}
Kensuke Onuma, Christos Faloutsos, and Jessica~K Hodgins.
\newblock Fmdistance: A fast and effective distance function for motion capture data.
\newblock \emph{Eurographics (Short Papers)}, 7\penalty0 (10), 2008.

\bibitem[Petrovich et~al.(2024)Petrovich, Litany, Iqbal, Black, Varol, Peng, and Rempe]{petrovich24stmc}
Mathis Petrovich, Or Litany, Umar Iqbal, Michael~J. Black, G{\"u}l Varol, Xue~Bin Peng, and Davis Rempe.
\newblock Multi-track timeline control for text-driven 3d human motion generation.
\newblock In \emph{CVPR Workshop on Human Motion Generation}, 2024.

\bibitem[Pinyoanuntapong et~al.(2024)Pinyoanuntapong, Wang, Lee, and Chen]{pinyoanuntapong2024mmm}
Ekkasit Pinyoanuntapong, Pu Wang, Minwoo Lee, and Chen Chen.
\newblock Mmm: Generative masked motion model.
\newblock In \emph{CVPR}, 2024.

\bibitem[Ponton et~al.(2023)Ponton, Yun, Aristidou, Andujar, and Pelechano]{ponton2023sparseposer}
Jose~Luis Ponton, Haoran Yun, Andreas Aristidou, Carlos Andujar, and Nuria Pelechano.
\newblock Sparseposer: Real-time full-body motion reconstruction from sparse data.
\newblock \emph{ACM Transactions on Graphics}, 2023.

\bibitem[Qian et~al.(2023)Qian, Urbanek, Hauptmann, and Won]{Qian_2023_ICCV}
Yijun Qian, Jack Urbanek, Alexander~G. Hauptmann, and Jungdam Won.
\newblock Breaking the limits of text-conditioned {3D} motion synthesis with elaborative descriptions.
\newblock In \emph{ICCV}, 2023.

\bibitem[Raffel et~al.(2020)Raffel, Shazeer, Roberts, Lee, Narang, Matena, Zhou, Li, and Liu]{2020t5}
Colin Raffel, Noam Shazeer, Adam Roberts, Katherine Lee, Sharan Narang, Michael Matena, Yanqi Zhou, Wei Li, and Peter~J. Liu.
\newblock Exploring the limits of transfer learning with a unified text-to-text transformer.
\newblock \emph{Journal of Machine Learning Research}, 21\penalty0 (140):\penalty0 1--67, 2020.

\bibitem[Rempe et~al.(2021)Rempe, Birdal, Hertzmann, Yang, Sridhar, and Guibas]{rempe2021humor}
Davis Rempe, Tolga Birdal, Aaron Hertzmann, Jimei Yang, Srinath Sridhar, and Leonidas~J. Guibas.
\newblock Humor: 3d human motion model for robust pose estimation.
\newblock In \emph{ICCV}, 2021.

\bibitem[Rombach et~al.(2022)Rombach, Blattmann, Lorenz, Esser, and Ommer]{Rombach_2022_CVPR}
Robin Rombach, Andreas Blattmann, Dominik Lorenz, Patrick Esser, and Bj\"orn Ommer.
\newblock High-resolution image synthesis with latent diffusion models.
\newblock In \emph{CVPR}, 2022.

\bibitem[S\'ar\'andi and Pons-Moll(2024)]{sarandi2024nlf}
Istv\'an S\'ar\'andi and Gerard Pons-Moll.
\newblock Neural localizer fields for continuous 3d human pose and shape estimation.
\newblock 2024.

\bibitem[Shen et~al.(2024)Shen, Pi, Xia, Cen, Peng, Hu, Bao, Hu, and Zhou]{shen2024world}
Zehong Shen, Huaijin Pi, Yan Xia, Zhi Cen, Sida Peng, Zechen Hu, Hujun Bao, Ruizhen Hu, and Xiaowei Zhou.
\newblock World-grounded human motion recovery via gravity-view coordinates.
\newblock In \emph{SIGGRAPH Asia 2024 Conference Papers}, pages 1--11, 2024.

\bibitem[Shin et~al.(2024)Shin, Kim, Halilaj, and Black]{shin2024wham}
Soyong Shin, Juyong Kim, Eni Halilaj, and Michael~J Black.
\newblock Wham: Reconstructing world-grounded humans with accurate 3d motion.
\newblock In \emph{CVPR}, 2024.

\bibitem[Shiobara and Murakami(2021)]{Shiobara21_wgan}
Ayumi Shiobara and Makoto Murakami.
\newblock Human motion generation using wasserstein {GAN}.
\newblock In \emph{International Conference on Digital Signal Processing ({ICDSP})}, 2021.

\bibitem[Siyao et~al.(2022)Siyao, Yu, Gu, Lin, Wang, Qian, Loy, and Liu]{siyao2022bailando}
Li Siyao, Weijiang Yu, Tianpei Gu, Chunze Lin, Quan Wang, Chen Qian, Chen~Change Loy, and Ziwei Liu.
\newblock Bailando: 3d dance generation by actor-critic gpt with choreographic memory.
\newblock In \emph{Proceedings of the IEEE/CVF Conference on Computer Vision and Pattern Recognition}, pages 11050--11059, 2022.

\bibitem[Song et~al.(2020)Song, Meng, and Ermon]{song2020denoising}
Jiaming Song, Chenlin Meng, and Stefano Ermon.
\newblock Denoising diffusion implicit models.
\newblock In \emph{International Conference on Learning Representations}, 2020.

\bibitem[Su et~al.(2024)Su, Ahmed, Lu, Pan, Bo, and Liu]{su2024roformer}
Jianlin Su, Murtadha Ahmed, Yu Lu, Shengfeng Pan, Wen Bo, and Yunfeng Liu.
\newblock Roformer: Enhanced transformer with rotary position embedding.
\newblock \emph{Neurocomputing}, 568:\penalty0 127063, 2024.

\bibitem[Sun et~al.(2022)Sun, Wang, Hu, Lai, Jin, and Hu]{sun2022_dancing}
Jiangxin Sun, Chunyu Wang, Huang Hu, Hanjiang Lai, Zhi Jin, and Jian-Fang Hu.
\newblock You never stop dancing: Non-freezing dance generation via bank-constrained manifold projection.
\newblock In \emph{NeurIPS}, 2022.

\bibitem[Sun et~al.(2023)Sun, Bao, Liu, Mei, and Black]{sun2023trace}
Yu Sun, Qian Bao, Wu Liu, Tao Mei, and Michael~J Black.
\newblock Trace: 5d temporal regression of avatars with dynamic cameras in 3d environments.
\newblock In \emph{CVPR}, 2023.

\bibitem[Tang et~al.(2018)Tang, Jia, and Mao]{tang2018_dance}
Taoran Tang, Jia Jia, and Hanyang Mao.
\newblock Dance with melody: An {LSTM}-autoencoder approach to music-oriented dance synthesis.
\newblock In \emph{ACM International Conference on Multimedia ({ACMMM})}, 2018.

\bibitem[Teed and Deng(2021)]{teed2021droid}
Zachary Teed and Jia Deng.
\newblock {DROID-SLAM: Deep Visual SLAM for Monocular, Stereo, and RGB-D Cameras}.
\newblock \emph{NeurIPs}, 2021.

\bibitem[Tevet et~al.(2023)Tevet, Raab, Gordon, Shafir, Cohen-Or, and Bermano]{tevet2023human}
Guy Tevet, Sigal Raab, Brian Gordon, Yonatan Shafir, Daniel Cohen-Or, and Amit~H Bermano.
\newblock Human motion diffusion model.
\newblock In \emph{ICLR}, 2023.

\bibitem[Tseng et~al.(2023)Tseng, Castellon, and Liu]{tseng2023edge}
Jonathan Tseng, Rodrigo Castellon, and Karen Liu.
\newblock Edge: Editable dance generation from music.
\newblock In \emph{Proceedings of the IEEE/CVF Conference on Computer Vision and Pattern Recognition}, pages 448--458, 2023.

\bibitem[Valle-Pérez et~al.(2021)Valle-Pérez, Henter, Beskow, Holzapfel, Oudeyer, and Alexanderson]{valleperez2021transflower}
Guillermo Valle-Pérez, Gustav~Eje Henter, Jonas Beskow, André Holzapfel, Pierre-Yves Oudeyer, and Simon Alexanderson.
\newblock Transflower: probabilistic autoregressive dance generation with multimodal attention.
\newblock \emph{ACM Transactions on Graphics ({TOG})}, 2021.

\bibitem[von Marcard et~al.(2018)von Marcard, Henschel, Black, Rosenhahn, and Pons-Moll]{vonMarcard2018}
Timo von Marcard, Roberto Henschel, Michael Black, Bodo Rosenhahn, and Gerard Pons-Moll.
\newblock Recovering accurate 3d human pose in the wild using imus and a moving camera.
\newblock In \emph{ECCV}, 2018.

\bibitem[Wang and Daniilidis(2023)]{refit}
Yufu Wang and Kostas Daniilidis.
\newblock Refit: Recurrent fitting network for 3d human recovery.
\newblock In \emph{Proceedings of the IEEE/CVF International Conference on Computer Vision}, pages 14644--14654, 2023.

\bibitem[Wang et~al.(2024)Wang, Wang, Liu, and Daniilidis]{wang2024tram}
Yufu Wang, Ziyun Wang, Lingjie Liu, and Kostas Daniilidis.
\newblock Tram: Global trajectory and motion of 3d humans from in-the-wild videos.
\newblock In \emph{European Conference on Computer Vision}, pages 467--487. Springer, 2024.

\bibitem[Wang et~al.(2022)Wang, Chen, Liu, Zhu, Liang, and Huang]{wang2022humanise}
Zan Wang, Yixin Chen, Tengyu Liu, Yixin Zhu, Wei Liang, and Siyuan Huang.
\newblock {HUMANISE}: Language-conditioned human motion generation in 3d scenes.
\newblock In \emph{Neural Information Processing Systems ({NeurIPS})}, 2022.

\bibitem[Xu et~al.(2023)Xu, Song, Wang, Su, Fang, Ding, Gan, Yan, Jin, Yang, Zeng, and Wu]{xu2023actformer}
Liang Xu, Ziyang Song, Dongliang Wang, Jing Su, Zhicheng Fang, Chenjing Ding, Weihao Gan, Yichao Yan, Xin Jin, Xiaokang Yang, Wenjun Zeng, and Wei Wu.
\newblock {ActFormer}: A gan-based transformer towards general action-conditioned 3d human motion generation.
\newblock In \emph{ICCV}, 2023.

\bibitem[Xu et~al.(2024)Xu, Hua, Lin, Liu, Ma, Yan, Jin, Yang, and Zeng]{xu2024motionbank}
Liang Xu, Shaoyang Hua, Zili Lin, Yifan Liu, Feipeng Ma, Yichao Yan, Xin Jin, Xiaokang Yang, and Wenjun Zeng.
\newblock Motionbank: A large-scale video motion benchmark with disentangled rule-based annotations, 2024.

\bibitem[Ye et~al.(2023)Ye, Pavlakos, Malik, and Kanazawa]{ye2023slahmr}
Vickie Ye, Georgios Pavlakos, Jitendra Malik, and Angjoo Kanazawa.
\newblock Decoupling human and camera motion from videos in the wild.
\newblock In \emph{CVPR}, 2023.

\bibitem[Yi et~al.(2024)Yi, Thies, Black, Peng, and Rempe]{yi2025generating}
Hongwei Yi, Justus Thies, Michael~J Black, Xue~Bin Peng, and Davis Rempe.
\newblock Generating human interaction motions in scenes with text control.
\newblock In \emph{European Conference on Computer Vision}, pages 246--263. Springer, 2024.

\bibitem[Yi et~al.(2021)Yi, Zhou, and Xu]{TransPoseSIGGRAPH2021}
Xinyu Yi, Yuxiao Zhou, and Feng Xu.
\newblock Transpose: Real-time 3d human translation and pose estimation with six inertial sensors.
\newblock \emph{ACM Transactions on Graphics}, 2021.

\bibitem[Yuan et~al.(2022)Yuan, Iqbal, Molchanov, Kitani, and Kautz]{yuan2022glamr}
Ye Yuan, Umar Iqbal, Pavlo Molchanov, Kris Kitani, and Jan Kautz.
\newblock Glamr: Global occlusion-aware human mesh recovery with dynamic cameras.
\newblock In \emph{CVPR}, 2022.

\bibitem[Zhang et~al.(2022)Zhang, Cai, Pan, Hong, Guo, Yang, and Liu]{zhang2022motiondiffuse}
Mingyuan Zhang, Zhongang Cai, Liang Pan, Fangzhou Hong, Xinying Guo, Lei Yang, and Ziwei Liu.
\newblock Motiondiffuse: Text-driven human motion generation with diffusion model.
\newblock \emph{arXiv preprint arXiv:2208.15001}, 2022.

\bibitem[Zhang et~al.(2023{\natexlab{a}})Zhang, Guo, Pan, Cai, Hong, Li, Yang, and Liu]{zhang2023remodiffuse}
Mingyuan Zhang, Xinying Guo, Liang Pan, Zhongang Cai, Fangzhou Hong, Huirong Li, Lei Yang, and Ziwei Liu.
\newblock Remodiffuse: Retrieval-augmented motion diffusion model.
\newblock In \emph{arXiv preprint arXiv:2304.01116}, 2023{\natexlab{a}}.

\bibitem[Zhang et~al.(2024{\natexlab{a}})Zhang, Jin, Gu, Hong, Cai, Huang, Zhang, Guo, Yang, He, et~al.]{zhang2024large}
Mingyuan Zhang, Daisheng Jin, Chenyang Gu, Fangzhou Hong, Zhongang Cai, Jingfang Huang, Chongzhi Zhang, Xinying Guo, Lei Yang, Ying He, et~al.
\newblock Large motion model for unified multi-modal motion generation.
\newblock In \emph{ECCV}, 2024{\natexlab{a}}.

\bibitem[Zhang et~al.(2023{\natexlab{b}})Zhang, Song, Huang, Chen, and yu~Liu]{zhange2023diffcollage}
Qinsheng Zhang, Jiaming Song, Xun Huang, Yongxin Chen, and Ming yu Liu.
\newblock Diffcollage: Parallel generation of large content with diffusion models.
\newblock In \emph{CVPR}, 2023{\natexlab{b}}.

\bibitem[Zhang et~al.(2024{\natexlab{b}})Zhang, Bhatnagar, Xu, Winkler, Kadlecek, Tang, and Bogo]{zhang2024rohm}
Siwei Zhang, Bharat~Lal Bhatnagar, Yuanlu Xu, Alexander Winkler, Petr Kadlecek, Siyu Tang, and Federica Bogo.
\newblock Rohm: Robust human motion reconstruction via diffusion.
\newblock In \emph{CVPR}, 2024{\natexlab{b}}.

\bibitem[Zhang et~al.(2025)Zhang, Bhatnagar, Starke, Petrov, Guzov, Dhamo, Pérez~Pellitero, and Pons-Moll]{zhang2024force}
Xiaohan Zhang, Bharat~Lal Bhatnagar, Sebastian Starke, Ilya~A. Petrov, Vladimir Guzov, Helisa Dhamo, Eduardo Pérez~Pellitero, and Gerard Pons-Moll.
\newblock Force: Dataset and method for intuitive physics guided human-object interaction.
\newblock In \emph{International Conference on 3D Vision (3DV)}, 2025.

\bibitem[Zhou et~al.(2024)Zhou, Dou, Cao, Liao, Wang, Wang, Liu, Komura, Wang, and Liu]{zhou2024emdm}
Wenyang Zhou, Zhiyang Dou, Zeyu Cao, Zhouyingcheng Liao, Jingbo Wang, Wenjia Wang, Yuan Liu, Taku Komura, Wenping Wang, and Lingjie Liu.
\newblock Emdm: Efficient motion diffusion model for fast and high-quality motion generation.
\newblock In \emph{European Conference on Computer Vision}, pages 18--38. Springer, 2024.

\bibitem[Zhou and Wang(2023)]{Zhou_2023_CVPR}
Zixiang Zhou and Baoyuan Wang.
\newblock Ude: A unified driving engine for human motion generation.
\newblock In \emph{CVPR}, 2023.

\bibitem[Zhu et~al.(2023{\natexlab{a}})Zhu, Liu, Liu, Qian, Liu, and Yu]{zhu2023taming}
Lingting Zhu, Xian Liu, Xuanyu Liu, Rui Qian, Ziwei Liu, and Lequan Yu.
\newblock Taming diffusion models for audio-driven co-speech gesture generation.
\newblock In \emph{CVPR}, 2023{\natexlab{a}}.

\bibitem[Zhu et~al.(2023{\natexlab{b}})Zhu, Ma, Ro, Ci, Zhang, Shi, Gao, Tian, and Wang]{zhu2023human}
Wentao Zhu, Xiaoxuan Ma, Dongwoo Ro, Hai Ci, Jinlu Zhang, Jiaxin Shi, Feng Gao, Qi Tian, and Yizhou Wang.
\newblock Human motion generation: A survey.
\newblock \emph{TPAMI}, 2023{\natexlab{b}}.

\end{thebibliography}
}

\appendix
\section{Implementation Details}
\label{sec:implementation}
\paragraph{Model Architecture.}
GENMO comprises 16 layers, each consisting of a ROPE-based Transformer block followed by a multi-text injection block. The ROPE-based Transformer block incorporates a LayerNorm, a ROPE attention layer with residual connections, and an MLP layer. Each attention unit features 8 attention heads to capture diverse motion patterns. The number of neurons in the MLP layer is $d_\text{mlp} = 1024$. The multi-text injection block maintains a similar architecture to the ROPE-based Transformer block, but replaces the standard attention with multi-text attention, which processes text embedding sequences to enrich and update the motion feature representations. The maximum self-attention window size is $W=120$.

\paragraph{Training Datasets.}
GENMO is trained from scratch on a diverse set of mixed motion datasets, including motion estimation datasets AMASS~\cite{AMASS:ICCV:2019}, BEDLAM~\cite{Black_CVPR_2023}, Human3.6M~\cite{h36m_pami}, 3DPW~\cite{vonMarcard2018}, music-to-dance dataset AIST++\cite{li2021ai}, and text-to-motion datasets HumanML3D~\cite{Guo_2022_CVPR} and Motion-X~\cite{lin2023motionx}. Since motion data in HumanML3D are represented in their own format, we convert them to SMPL parameters with inverse kinematics~\cite{li2020hybrik} for training. For AMASS data lacking video, music, or text inputs, we follow~\cite{shin2024wham,shen2024world} to simulate static and dynamic camera trajectories and project 3D motions to 2D keypoints as input conditions. The simulated camera trajectories are also used as input conditions during training. Although AMASS and HumanML3D share some motion sequences, we treat them as independent datasets.

For Motion-X, we only utilize its 2D keypoints and text descriptions due to noisy 3D ground truth. When training with BEDLAM and Human3.6M datasets, we use video frames and 2D keypoints as conditioning inputs, with global 3D motions serving as target outputs. For the 3DPW dataset, video frames and 2D keypoints are used as conditions; however, since 3DPW provides only local 3D motions, we implement a strategy analogous to $\mathcal{L}_\text{gen-2D}$: we first generate pseudo-clean global human trajectories from the estimation mode, then utilize these to produce noisy motions for training the generation mode, with loss computation restricted to local poses. For AIST++, training incorporates video frames, 2D keypoints, and music as conditions. Regarding the camera condition, we utilize ground-truth camera trajectories as the input condition for datasets that either provide such trajectories or feature static cameras; for datasets lacking labeled camera trajectories, we employ DROID-SLAM~\cite{teed2021droid} to generate camera trajectories as input conditions during training. We train a single unified model on this comprehensive collection of datasets, enabling evaluation across diverse motion-related tasks.

\paragraph{Condition Processing.}
For video conditions, we employ a frozen encoder from TRAM~\cite{wang2024tram}, whereas for AMASS data lacking video inputs, we utilize zero vectors as placeholders. The 2D keypoint conditions undergo normalization to the range $[-1, 1]$ based on their bounding boxes, which are further normalized by the focal length of their corresponding video conditions. For music processing, we extract features using the music encoder from EDGE~\cite{tseng2023edge}, while camera parameters are formulated as the camera-to-world transformation and derived from input videos via DROID-SLAM~\cite{teed2021droid}. Textual descriptions are encoded through the T5 encoder architecture~\cite{2020t5}.

\paragraph{Training Details.}
During training, we employ data augmentation techniques on the 2D keypoints, including random masking and Gaussian noise perturbation to enhance model robustness. To further improve model robustness, we implement random masking of input conditions throughout the training process. We configure the sequence length to $N=120$ for training, while maintaining support for variable sequence lengths during inference. The model is trained from scratch for $500$ epochs using the AdamW optimizer~\cite{loshchilov2017fixing}, with a mini-batch size of $128$ per GPU distributed across $2$ A100 GPUs.

\section{Evaluation Settings for Music-to-Dance Generation}
We evaluate the music-to-dance generation capabilities of GENMO on the AIST++~\cite{li2021ai} dataset. The same one-in-all checkpoint is employed for evaluation as used in all other tasks. Following established protocols~\cite{li2021ai,tseng2023edge}, our evaluation encompasses four key aspects: motion quality, generation diversity, physical plausibility, and motion-music correlation. 

For motion quality and generation diversity assessment, we compute the Fréchet Inception Distance (FID)~\cite{heusel2017gans} and the average feature distance of generated motions using both kinetic features~\cite{onuma2008fmdistance} (denoted as ``k'') and geometric features~\cite{muller2005efficient} (denoted as ``g'') in accordance with Li \etal~\cite{li2021ai}. 

To evaluate physical plausibility, we employ two metrics: Mean Per Joint Position Error (MPJPE) and Procrustes-aligned MPJPE (PA-MPJPE). Additionally, we calculate the Physical Foot Contact score (PFC) as proposed by Tseng \etal~\cite{tseng2023edge}.

For quantifying motion-music correlation, we utilize the Beat Alignment Score (BAS) following the methodology of Li \etal~\cite{siyao2022bailando}. This metric effectively measures the synchronization between musical beats and motion transitions by calculating the average temporal distance between each kinematic beat and its nearest musical beat.

\section{Evaluation Settings for Text-to-Motion Generation}
For evaluating text-to-motion generation on HumanML3D~\cite{Guo_2022_CVPR}, we utilize the pre-trained text and motion encoders from~\cite{Guo_2022_CVPR} after converting our motion representation to the HumanML3D format. This conversion process involves first recovering the SMPL parameters from our raw representation and subsequently deriving the HumanML3D-format representation as described in~\cite{Guo_2022_CVPR}, employing the neutral gender SMPL model. Consistent with established evaluation protocols, we report the variance across five different inference trials on HumanML3D. The same one-in-all checkpoint is employed for evaluation as used in all other tasks.

It is important to note that the conversion from SMPL to HumanML3D format introduces some degradation in motion quality, as the predicted SMPL bone lengths do not precisely match the HumanML3D skeleton, resulting in artifacts such as foot skating. To address this limitation and provide a more comprehensive evaluation, we additionally report the Fréchet Inception Distance (FID) and Diversity metrics using both kinetic features~\cite{onuma2008fmdistance} (denoted as ``k'') and geometric features~\cite{muller2005efficient} (denoted as ``g'') based on 24 keypoints. Since the SMPL model and the HumanML3D skeleton share an identical joint order, this approach enables direct comparison of GENMO's motion quality with state-of-the-art methods using keypoint-based metrics.

For the evaluation on Motion-X~\cite{lin2023motionx}, we implemented our own text and motion encoders, as the original encoders were provided by ~\cite{lin2023motionx} and their implementation details were not disclosed in the literature. Unlike HumanML3D, Motion-X text prompts lack frame-based keywords, necessitating a different approach to text encoding. We employed a pre-trained CLIP language model with its corresponding tokenizer to process the raw text prompts, generating embeddings with a dimension of 512, consistent with the representation used in~\cite{Guo_2022_CVPR}. The same one-in-all checkpoint is employed for evaluation as used in all other tasks. For evaluation purposes on the Motion-X dataset, we utilized these trained encoders with frozen weights to ensure consistent and comparable feature extraction across all test samples.

\section{Evaluation Settings for Motion In-betweening}
For motion in-betweening evaluation, we adopt the methodology established in prior diffusion-based approaches~\cite{tevet2023human}, wherein the noisy motion is overwritten with desired poses at specified keyframes prior to each denoising step. The same one-in-all checkpoint is employed for evaluation as used in all other tasks. Due to the constraints of our feature representation, which lacks global root information, we only overwrite the local body poses and global root orientation for the keyframes. We evaluate our approach on both the HumanML3D and Motion-X test sets under two experimental conditions: sampling either 2 or 5 keyframes from each test motion. For Motion-X, we utilize the reconstructed 3D motion as described in~\cite{lin2023motionx}. Additionally, we incorporate textual descriptions from these datasets as conditioning input. To account for the generative diversity of our model, we sample $N=10$ different initial noise vectors for each test motion and execute the diffusion process with 50 denoising steps. For evaluation metrics, we report the minimum values among these diverse samples, which effectively captures the best performance achievable by our generative approach.

\begin{table}[t]
    \centering
    \caption{
        \textbf{Benchmark of Human Motion Generation.} Motion quality is evaluated on the 3DPW-XOCC~\cite{li2023niki} dataset.
    }
    \label{tab:quant_3dpw_xocc}
    \setlength{\tabcolsep}{5pt} %
    \renewcommand{\arraystretch}{1.2} %
    \resizebox{\columnwidth}{!}{ %
        \begin{tabular}{l|cccc}
            \toprule
            \textbf{Methods} & MPJPE~$\uparrow$ & PA-MPJPE~$\downarrow$ & PVE~$\downarrow$ & ACCEL~$\rightarrow$ \\
            \midrule
            HybrIK~\cite{li2020hybrik} & 148.3 & 98.7 & 164.5 & 108.6 \\
            PARE~\cite{pare2021kocabas} & 114.2 & 67.7 & 133.0 & 90.7 \\
            PARE~\cite{pare2021kocabas} + VIBE~\cite{kocabas2020vibe} & 97.3 & 60.2 & 114.9 & 18.3 \\
            NIKI (frame-based) ~\cite{li2023niki} & 110.7 & 60.5 & 128.6 & 74.4 \\
            NIKI (temporal)~\cite{li2023niki} & 88.9 & 52.1 & 98.0 & 17.3 \\
            \midrule
            Ours (Regression-only) & {89.0} & {50.2} & {103.8} & {21.1} \\
            Ours & \textbf{76.2} & \textbf{48.4} & \textbf{94.2} & \textbf{17.1} \\
            \bottomrule
        \end{tabular}
    }
\end{table}

\section{Evaluation on Occlusion-Specific Benchmark}
\label{sec:eval_3dpw_xocc}

To evaluate the efficacy of generative priors in enhancing motion estimation robustness, we conducted comprehensive experiments on the 3DPW-XOCC benchmark~\cite{li2023niki}. This benchmark specifically evaluates 3D human pose estimation under challenging conditions of extreme occlusion and truncation, simulated through strategic placement of random occlusion patches and frame truncations. As evidenced in Table~\ref{tab:quant_3dpw_xocc}, GENMO demonstrates superior performance compared to state-of-the-art human motion estimation methods, including those explicitly designed to handle occlusions. Notably, our ablation study reveals that a variant of our model trained without generative tasks exhibits worse performance compared to the complete GENMO model. These findings substantiate that the generative priors incorporated within GENMO significantly enhance the plausibility and accuracy of estimated human motions under visually challenging scenarios, thereby underscoring the practical utility of our approach in real-world applications where occlusions frequently occur.

\section{Additional Related Work} 
\subsection{Generative Priors for Estimation}
Recent advances in computer vision have demonstrated the efficacy of leveraging generative priors from large-scale image models, such as StableDiffusion~\cite{Rombach_2022_CVPR}, for various estimation tasks. These approaches fine-tune diffusion-based generative models to predict geometric and semantic properties, including depth maps, surface normals, and semantic segmentation~\cite{fu2024geowizard, ke2023repurposing, martingarcia2024diffusione2eft}. By repurposing the rich latent representations encoded in pre-trained generative models, these methods achieve substantial improvements in estimation accuracy across diverse visual understanding tasks. Nevertheless, a significant limitation of these approaches is their tendency to sacrifice the inherent generative capabilities of the original models, as they predominantly focus on deterministic estimation outcomes rather than maintaining the ability to produce diverse outputs.

Our work fundamentally diverges from these approaches by introducing a unified framework that seamlessly integrates motion generation and estimation within a single coherent model. In contrast to previous methods that compromise generative capabilities during the fine-tuning process, our framework maintains both the stochastic diversity essential for high-quality generation and the deterministic precision required for accurate estimation. This dual capability represents a significant advancement in leveraging generative priors for human motion understanding.

\end{document}